\documentclass[traditabstract]{aa} 
\newcommand{\Msun}{\mbox{\,$M_{\odot}$\/}}          
\newcommand{\Rsolar}{\mbox{\,$R_{\odot}$\/}}          
\newcommand{\Mjup}{\mbox{\,$M_{\rm Jup}$\/}}          
\newcommand{\teff}{\it{T}\rm$_{\rm{eff}}$}
\newcommand{\logg}{log\,\it{g}\rm}

\usepackage{lscape}
\usepackage{txfonts}
\usepackage{natbib}
\usepackage{graphicx}

\begin{document}
   \title{Spectroscopy across the brown dwarf/planetary mass boundary - I. \\
     Near-infrared JHK spectra\thanks{based on observations obtained at the Paranal Observatory
    , Chile for ESO programs 279.C-5010(A), 080.C-0590(A), 077.C-0264(A), 078.C-0800(B), \& 078.C-0800(A)}}

   \author{J. Patience\inst{1}, R. R. King\inst{1}, R. J. De Rosa \inst{1},
     A. Vigan \inst{1}, S. Witte \inst{2}, E. Rice \inst{3}, Ch. Helling \inst{4}, 
          \and
          P. Hauschildt \inst{2} 
          }

   \institute{Astrophysics Group, School of Physics, University of Exeter, Exeter, EX4 4QL,
     UK\\
\email{patience@astro.ex.ac.uk, rob@astro.ex.ac.uk, derosa@astro.ex.ac.uk, arthur@astro.ex.ac.uk}
              \and
             Hamburger Sternwarte, Gojenbergsweg 112, 21029 Hamburg,
             Germany\\
\email{switte@hs.uni-hamburg.de, yeti@hs.uni-hamburg.de}
             \and
             Department of Astrophysics, American Museum of Natural History, 79th Street and Central Park West, New York, NY 10024, USA; Current address: Department of Engineering Science \& Physics, College of Staten Island, 2800 Victory Blvd, Staten Island, NY 10314, USA\\
\email{erice@amnh.org}
             \and
             SUPA, School of Physics \& Astronomy, University of
             St. Andrews, North Haugh, St. Andrews, KY16 9SS, UK\\
\email{ch80@st-andrews.ac.uk}
             }


   \date{Received Sep. 9, 2011; accepted Jan. 5, 2012}

 
  \abstract
  { With a uniform VLT SINFONI data set of nine targets, we have developed an empirical
  grid of J,H,K spectra of the atmospheres of objects estimated to have very low substellar
  masses of $\sim$5-20\Mjup\ and young ages ranging from $\sim$1-50 Myr. Most of the targets
  are companions, objects which are especially valuable for comparison with atmosphere and
  evolutionary models, as they present rare cases in which the age is accurately known
  from the primary. Based on the youth of the sample, all objects are expected to have low
  surface gravity, and this study investigates the critical early phases of the evolution of
  substellar objects. The spectra are compared with grids of five different theoretical
  atmosphere models. This analysis represents the first systematic model comparison with
  infrared spectra of young brown dwarfs. The fits to the full JHK spectra of each object
  result in a range of best fit effective temperatures of $\pm$150-300K whether or not the
  full model grid or a subset restricted to lower log(g) values is used. This effective
  temperature range is significantly larger than the uncertainty typically assigned when
  using a single model grid. Fits to a single wavelength band can vary by up to 1000K using
  the different model grids. Since the overall shape of these spectra is governed more by
  the temperature than surface gravity, unconstrained model fits did not find matches with
  low surface gravity or a trend in log(g) with age. This suggests that empirical comparison
  with spectra of unambiguously young objects targets (such as those presented here) may be
  the most reliable method to search for indications of low surface gravity and youth. Based
  on comparison with previous observations, the SINFONI spectra represent a second epoch for
  the targets 2M0141 and DH Tau B, and the combined data show no variations in the spectral
  morphology over time. The analysis of two other targets, AB Pic B and CT Cha B, suggests
  that these objects may have lower temperatures, and consequently lower masses, than
  previously estimated.}


 \authorrunning{Patience, King, De Rosa, Vigan, Witte, Rice, Helling and Hauschildt}
 \titlerunning{Near-IR spectra across the brown dwarf/planet boundary}
   \maketitle
%

\section{Introduction}

Since the discovery of the first substellar companions GD 165B \citep{Becklin:1988} and  Gl
229B \citep{Nakajima:1995}, there have been many surveys for additional low mass companions,
but a very limited number have been discovered \citep[e.g.,][]{McCarthy:2004, Metchev:2006},
particularly at the lowest masses \citep{Zuckerman:2009}. Substellar objects that are
companions to nearby stars and brown dwarfs are especially valuable for comparison with
atmosphere and evolutionary models, as they present rare cases in which the brown
dwarf age is inferred by the accurate age of the primary. Recent models
\citep[e.g.,][]{Fortney:2008} have indicated the importance initial conditions may have in
determining the flux from young objects, making confirmed young brown dwarfs important to
gauge the effects of accretion history and atmospheric processes at low surface gravity.

The effects of low surface gravity on the infrared spectra of brown dwarfs have been
investigated with isolated members of star-forming regions and young clusters. In the
J-band, the alkali metal lines of K I and Na I and the FeH absorption were found to have
lower equivalent widths in young brown dwarfs relative to field objects of the same spectral
type in an initial sample of three young objects \citep{McGovern:2004} and a larger sample
of 23 brown dwarfs in Upper Sco \citep{Lodieu:2008}. Lower pressures in the atmospheres of
the young objects cause a lower amount of recombination of K II and Na II to form K I and Na
I which, along with pressure-broadening \citep{Rice:2010} and other processes, impacts
the line strengths relative to the field objects, resulting in weaker lines. An index based
on the Na I doublet was defined to assess surface gravity in the youngest objects
\citep{Allers:2007}. The triangular shape of the H-band was noted in the spectra of low
luminosity $\sigma$ Orionis members \citep{Lucas:2001} and associated with a decrease in
H$_{2}$ collision induced absorption due to low gravity in a young field object
\citep{Kirkpatrick:2006}. \citet{Rice:2011} demonstrate how the relative opacities from
CIA-H$_2$ absorption and H$_2$O at different gravities shape the
spectra of young objects. The
CO bandhead shape in the K-band has also been linked with lower surface gravity due to the
sensitivity of the absorption to microturbulence, and has been used to differentiate dwarf
and giant stars \citep{Kleinmann:1986}.

\begin{table*}
\caption{Sample}             
\label{table:sample}      
\renewcommand{\footnoterule}{}  
\centering          
\begin{tabular}{l c c c c c c c c c c}     
\hline\hline       
                      
Source & Primary & Comp. & Age & $M_{\rm{prim}}$ & $M_{\rm{sec}}$ &
D  & Sep. & Membership & A$_{J}$ & Refs.  \\ 
& & & (Myr) & (\Msun) & (\Mjup) & (pc) & (AU) &  & & \\
\hline                    
    GQ Lup B 	& K7   & L1.5   	& 1     & 0.7   		& 24$\pm$12 		& 150$\pm$20  	& 100 & Lupus SFR 			& 0.1$\pm$0.2 	& 1-5 \\
  	DH Tau B  	& M0.5 & L2     	& 1 	& 0.33  		& 11$\pm{10\atop3}$ & 140$\pm$15 	& 330 & Taurus SFR 			& 0.3$\pm$0.3 	& 6,2  \\ 
	CT Cha  B  	& K7   & $\ge$M8	& 2 	& 0.7 			& 17$\pm$6  		& 165$\pm$30  	& 440 & Chameleon SFR 		& 0.4$\pm$0.2 	&  7-8 \\
\\
    2M1207 A\&B & M8   & L5     	& 8     & 25$\pm$5\Mjup & 6-10  			& 52.4$\pm$1.1 	& 46  & TW Hydra Assoc. 	& -- 			& 9-12 \\   
    TWA 5 B  	& M1.5 & M8/8.5 	& 8     & 0.40  		& $\sim$20 				& 44$\pm$4   	& 98  & TW Hydra Assoc. 	& -- 			& 13-15 \\
\\
    AB Pic B  	& K2   & L0-L1     	& 30    & 0.84  		& 13-14 				& 47.3$\pm$1.8 	& 248 & Tuc-Hor Assoc. 		& -- 			& 16-17 \\
    GSC08047 B 	& K3   & M9.5     	& 30 	& 0.80  		& 25$\pm$10 		& $\sim$85    	& 279 & Tuc-Hor Assoc. 		& -- 			& 18-19 \\
\\
    2M0141  	& L0   & --      	& 1-50  & 15$\pm$10\Mjup& -- 				& $\sim$35  	& --  & Tuc-Hor/$\beta$ Pic?& -- 			& 20 \\

\hline      
\end{tabular}
\\{\it References: } (1) \citet{Neuhauser:2005}  (2) \citet{Luhman:2006} 
(3) \citet{Marois:2007} (4) \citet{Franco:2002} (5) \citet{Batalha:2001} (6) \citet{Itoh:2005} 
(7) \citet{Schmidt:2008} (8) \citet{Luhman:2004} (9) \citet{Gizis:2002} 
(10) \citet{Chauvin:2004} (11) \citet{Chauvin:2005b} (12) \citet{Mohanty:2007}
(13) \citet{Lowrance:1999} (14) \citet{Webb:1999} (15) \citet{Konopacky:2007} 
(16) \citet{Chauvin:2005c} (17) \citet{Bonnefoy:2010} (18) \citet{Chauvin:2005a}
(19) \citet{Neuhauser:2004} (20) \citet{Kirkpatrick:2006}
\end{table*}

Given the intrinsically low luminosities and temperatures of substellar objects, infrared
observations provide higher signal-to-noise spectra than optical measurements, making this
wavelength range important for the characterization of the physical properties of brown
dwarfs. For example,  J-band spectra taken at two resolutions of a sample of M6-M9 objects
including 15 brown dwarfs with ages $\sim$1-10 Myr were compared with a grid of model
atmospheres to infer a number of properties -- effective temperatures, surface gravities,
radial velocities and projected rotational velocities \citep{Rice:2010}. 

This paper presents the combined flux-calibrated J,H,K spectra for a set of young (2-50
Myr), low mass (5-25\Mjup) substellar objects and compares the observations with a suite of
synthetic spectra developed from five different theoretical model atmosphere simulations.
The range of physical properties inferred from the different models is discussed, along with
the discrepancies obtained from fits over portions of the J,H,K wavelength range rather than
the full combination. The spectral shape is the focus of this paper, and the investigation
of spectral lines and longer wavelengths covering the L-band are the subject of a
forthcoming paper (King et al. 2011, in prep.).

\section{Sample and Summary of Previous Observations}

\begin{figure}
   \centering
   \resizebox{\hsize}{!}{\includegraphics{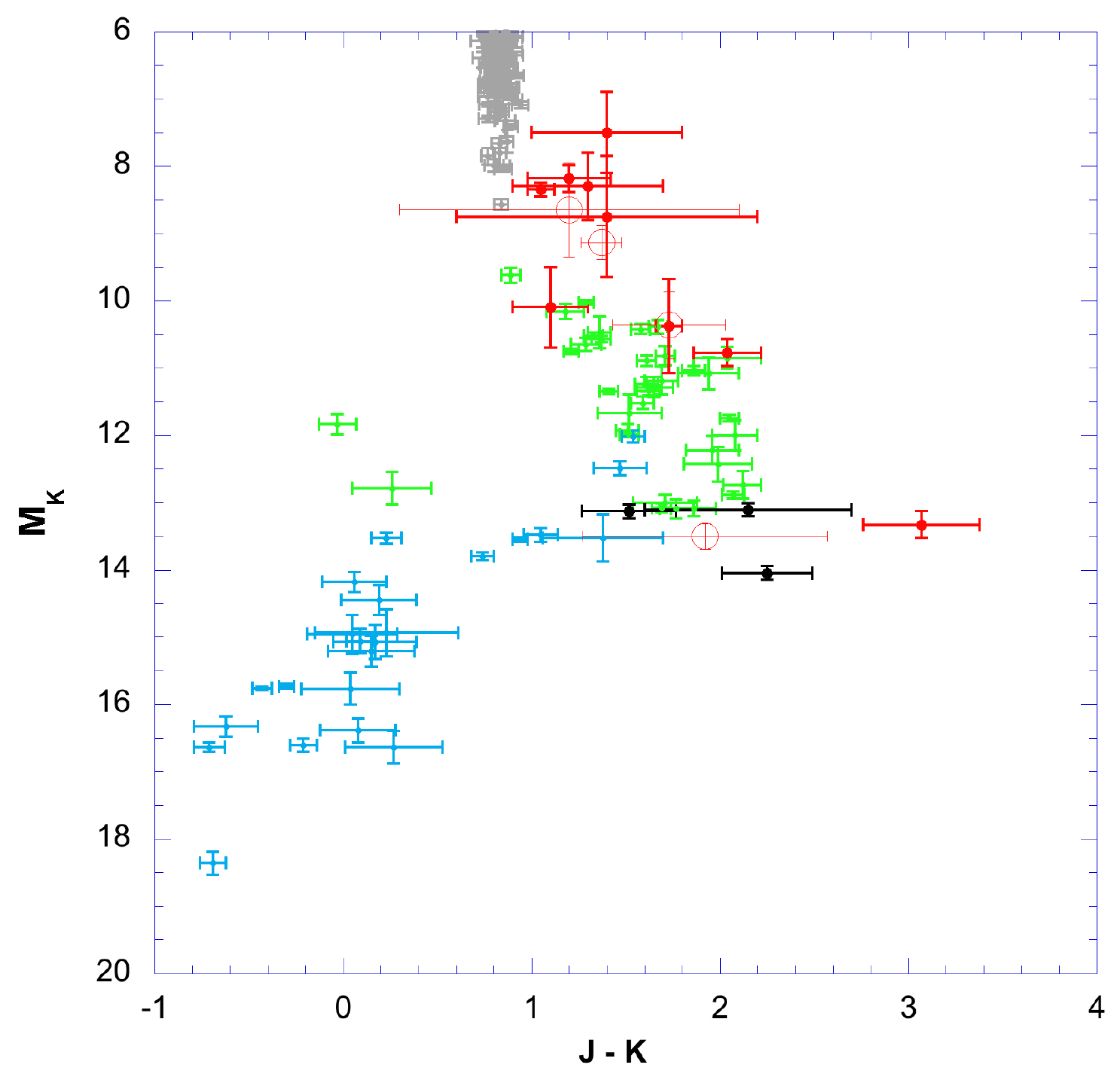}}
      \caption{Color-magnitude diagram of Field M (grey), L (green), T
        (blue) objects, the sample targets (filled red) and other
        young companions (unfilled red), and the HR 8799 planets with
        measured J and K magnitudes (black). The field data were taken
        from the DwarfArchive.org and restricted
        to objects with parallax uncertainty $<$10\% and color
        uncertainty $<$0.5 mag. The values used for the targets are given in Tables \ref{table:sample}
        and \ref{table:phot}. For the young companions CHXR 73B \citep{Luhman:2006},
        1RXS J1609B \citep{Lafreniere:2008}, GSC 06214 B \citep{Ireland:2011}
        and HD 203030B \citep{Metchev:2006} that are
        plotted on the figure, but not included in this study, the
        data were taken from the discovery papers. }
         \label{CMD}
   \end{figure}

The sample consists of 9 objects estimated to have very low masses of $\sim$5-20 $M_{Jup}$
and young ages ranging from $\sim$1-50 Myr, placing them at the boundary of the brown dwarf
and planetary mass regimes. The sample covers three age bins, with three $\sim$1-3 Myr
old members of star-forming regions, three members of the $\sim$8 Myr TW Hydra Association,
and three targets with ages of $\sim$30-50 Myr. Eight of the targets are members of binary
systems discovered with adaptive optics imaging surveys, and one object is an isolated young
brown dwarf. The youngest objects may still be surrounded by material from their formation
environment and have the possibility of accretion and disks, while the older objects likely
present bare photospheres. All are expected to have low surface gravity. The distinct
position (dereddened, when necessary) on the color-magnitude diagram of this young, low
surface gravity sample compared to field objects is shown in Figure \ref{CMD}; the trend is
as predicted in theoretical evolutionary models \citep[e.g.,][]{Saumon:2008}. The targets
show redder colors than field L and T dwarfs, analogous to the imaged planets orbiting HR
8799. For all members of the sample, there are archive infrared J,H,K spectra from the VLT
integral field spectrograph SINFONI, forming a uniform data set with which it is possible to
construct an empirical grid of the atmospheres of substellar objects covering the critical
early phases of their evolution.

\begin{table}
\caption{Photometry}             
\label{table:phot}      
\centering          
\begin{tabular}{l l c c c}     
\hline\hline       
                      
\multicolumn{2}{c}{Source} & J & H & K \\ 
\hline                    
CT Cha & A & 9.72 $\pm$ 0.02  & 8.94 $\pm$ 0.05  & 8.66  $\pm$ 0.02 \\
       & B & 16.6 $\pm$ 0.3   & ...              & 14.9 $\pm$ 0.3    \\
DH Tau & A & 9.77 $\pm$ 0.02  & 8.82 $\pm$ 0.03  & 8.18 $\pm$ 0.03  \\
       & B & 15.71 $\pm$ 0.05 & 14.96 $\pm$ 0.04 & 14.19 $\pm$ 0.02
          \\
GQ Lup & A & 8.69 $\pm$ 0.04  & 7.70 $\pm$ 0.03  & 7.10 $\pm$ 0.02  \\
       & B & 14.90 $\pm$ 0.11 & ...              & 13.34 $\pm$ 0.13  \\
\\
 
2M1207 & A & 13.00 $\pm$ 0.03 & 12.39 $\pm$ 0.03 & 11.95 $\pm$ 0.03  \\  
       & B & 20.0 $\pm$ 0.2   & 18.09 $\pm$ 0.21 & 16.93 $\pm$ 0.11   \\
TWA 5  & Aa  &  8.40 $\pm$ 0.07 & 7.69 $\pm$ 0.04  & 7.39 $\pm$ 0.04 \\
       & Ab  &  8.5 $\pm$ 0.2   & 7.79 $\pm$ 0.05  & 7.62 $\pm$ 0.08  \\
       & B  & 12.6 $\pm$ 0.2   & 12.14 $\pm$ 0.06 & 11.4 $\pm$ 0.2   \\
\\
  
AB Pic & A   & 7.58 $\pm$ 0.03  & 7.09 $\pm$ 0.03  & 6.98 $\pm$ 0.03 \\
       & B   & 16.18 $\pm$ 0.10 & 14.69 $\pm$ 0.10 & 14.14 $\pm$ 0.08 \\
GSC08047 & A & 9.06 $\pm$ 0.03  & 8.53 $\pm$ 0.06  & 8.41 $\pm$ 0.03  \\ 
         & B & 15.9 $\pm$ 0.1   & 15.45 $\pm$ 0.02 & 14.8 $\pm$ 0.1  \\
\\
 
   2M0141 &  & 14.83 $\pm$ 0.04 & 13.88 $\pm$ 0.03 & 13.10 $\pm$ 0.03 \\

\hline                  
\end{tabular}
\end{table}

\subsection{Star-forming region systems}

High angular resolution imaging surveys targeting members of the nearest star-forming
regions have identified a number of very young ($\sim$1-2 Myr) wide orbit ($>$100 AU) low
mass brown dwarf companions. The pre-Main Sequence sample considered for this paper are the
three companions with archive SINFONI J,H,K spectra: GQ Lup B \citep{Neuhauser:2005}, DH Tau
B \citep{Itoh:2005}, and CT Cha B \citep{Schmidt:2008}. The closest companion is GQ Lup B,
with a separation of 0$\farcs$7, or $\sim$105 AU at a distance of 150 $\pm$ 20 pc
\citep{Franco:2002}, and has been the subject of considerable prior investigation into the
inferred companion mass and evidence for ongoing accretion. Based on a comparison of
theoretical models with photometry and K-band spectroscopy, the mass was initially estimated
as 1-42\Mjup\ \citep{Neuhauser:2005}, while subsequent analysis of a larger wavelength range
\citep[e.g.,][]{Marois:2007} revised the mass range to 10-20\Mjup. In Table
\ref{table:sample}, the mass is listed as 24 $\pm$ 12 M$_{Jup}$ \citep{Luhman:2006}, since
this estimate applied the same approach to both GQ Lup B and another target DH Tau B. The
SINFONI spectra reported in \citet{Seifahrt:2007} suggested evidence for active accretion
from the Pa $\beta$ emission line, while independent J,H,K spectra from OSIRIS
\citep{McElwain:2007} and NIFS \citep{Lavigne:2009} do not confirm the line, which may
have resulted from contamination from the primary.The SINFONI spectrum of GQ Lup is
included in this paper for completeness.

The companion DH Tau B \citep{Itoh:2005} is in a wider orbit, at 2$\farcs$3 (330 AU) from
the primary, making contamination from the host star unlikely. The estimated mass for the
companion, 11$^{+10}_{-3}$\Mjup, listed in Table \ref{table:sample} is based on the same
model comparison applied to GQ Lup B \citep{Luhman:2006} and within the 5-40\Mjup\ range
reported in \citet{Itoh:2008}. The previous CISCO J,H,K spectrum (R$\sim$440) showed
absorption from H$_2$O bands indicating a low effective temperature, and no evidence of
emission lines from accretion was present \citep{Itoh:2005}. The SINFONI data presented in
this paper were taken at higher resolution (R$\sim$1500) than the CISCO spectrum and at a
different epoch, enabling a search for variability.

The widest young companion CT Cha B is 2$\farcs$7 (440 AU) from the primary
\citep{Schmidt:2008}. Based on fitting the SINFONI J,H,K spectra with Drift-PHOENIX models
to determine the temperature and photometry to estimate the luminosity, the companion mass
was estimated to be 17 $\pm$ 6\Mjup\ \citep{Schmidt:2008} from comparison with evolutionary
models \citep{Baraffe:2003,Chabrier:2000,Burrows:1997}. The spectrum shows Pa $\beta$
emission, an indication of ongoing accretion, though another accretion diagnostic Br
$\gamma$ is not seen \citep{Schmidt:2008}. In this paper, we reconsider the SINFONI spectra
in comparison with a larger set of models and with a different treatment of extinction.

\begin{table*}
\caption{SINFONI Observations}             
\label{table:obs}      
\centering          
\begin{tabular}{l c c c c c l l c}     
\hline\hline       
Source & Grating & Scale & NDIT & DIT & $\#$exp.  & Dates & ESO
Project ID &  PI \\ 
\hline  
   CT Cha B    & J   & 100mas & 2 & 138.5s & 6  & 18May07 &
   279.C-5010(A) &  Schmidt \\               
   CT Cha B    & H+K & 100mas & 2 & 138.5s & 6  & 16May07 & 279.C-5010(A) &  Schmidt\\
   DH Tau B    & J   & 25mas  & 1 & 300s   & 32 & 18Dec07 & 080.C-0590(A) &  Rojo\\
   DH Tau B    & H+K & 25mas  & 1 & 300s   & 8  & 22Oct07 & 080.C-0590(A) &  Rojo\\
   GQ Lup B    & J   & 25mas  & 1 & 300s   & 27 & 18Sep06 & 077.C-0264(A)  &   Neuh\"auser\\
   GQ Lup B    & H   & 25mas  & 1 & 300s   & 11 & 24Ap06  & 077.C-0264(A)  &    Neuh\"auser \\
   GQ Lup B    & K   & 25mas  & 1 & 300s   & 8  & 17Sep05  & 077.C-0264(A)  &  Neuh\"auser \\
\\

   TWA 5B      & J   & 25mas  & 1 & 120s   & 6  & 12Dec07  & 080.C-0590(A)  &   Rojo \\
   TWA 5B      & H+K & 25mas  & 2 & 60s    & 7  & 13Dec07  & 080.C-0590(A)  &    Rojo  \\
   2M1207A+B   & J   & 100mas & 1 & 600s   & 22 & 25Feb07  & 078.C-0800(B)  &   Thatte   \\
   2M1207A+B   & H+K & 100mas & 1 & 300s   & 24 & 28Jan07, 07Feb07  & 078.C-0800(B)  & Thatte  \\
\\

   AB Pic B    & J   & 25mas  & 1 & 300s   & 40 & 5-7Dec07 & {\bf 080.C-0590(A)}  &   Rojo  \\
   AB Pic B    & H+K & 25mas  & 1 & 300s   & 8  & 11Nov07  & {\bf 080.C-0590(A)}  &  Rojo   \\
   GSC 08047 B & J   & 25mas  & 1 & 300s   & 24 & 6\&18Jan08 & 080.C-0590(A) &   Rojo \\
   GSC 08047 B & H+K & 25mas  & 1 & 300s   & 27 & 5\&9Jan08 & 080.C-0590(A) &  Rojo \\
\\

   2M0141     & J   & 250 mas & 1 & 300 & 4 & 19Oct06 & 078.C-0800(A)
   &   Thatte \\
2M0141     & H+K   & 250 mas & 1 & 60s & 4 & 19Oct06 & 078.C-0800(A)
   &   Thatte \\
\hline                  
\end{tabular}
\end{table*}

\subsection{TW Hydra Association systems}

Two members of the TW Hydra Association -- TWA 5 and 2M1207 -- include three substellar
objects, with both late-M and mid-L spectral types. TWA 5B was one of the earliest brown
dwarf companions identified \citep{Lowrance:1999,Webb:1999} and is located 2$\farcs$0
($\sim$100 AU) from a close pair of early M-stars with a $\sim$5 yr orbit
\citep{Konopacky:2007}. Although the photometry of TWA 5B has been explored
\citep{Neuhauser:2000}, the H- and K-band spectra have been published separately
\citep{Neuhauser:2009}. The second TW Hydra member, 2M1207, is composed of two substellar
objects \citep{Chauvin:2004,Chauvin:2005b} -- the primary is an M8 brown dwarf
\citep{Gizis:2002} similar to TWA 5B and the secondary is a L-type planetary mass object
\citep{Chauvin:2004,Mohanty:2007}. The complete J,H,K spectrum of 2M1207B has been published
\citep{Patience:2010}, but not the corresponding spectrum of the primary.  When compared
with DUSTY models, the companion spectrum presents a spectral shape consistent with
temperatures higher than expected from the luminosity \citep{Mohanty:2007, Patience:2010}.
The discrepancy between spectral shape and flux level was initially explained by obscuration
from a grey disk \citep{Mohanty:2007}, and then updated to include models incorporating an
improved treatment of clouds \citep{Skemer:2011} and both clouds and non-equilibrium
chemistry \citep{Barman:2011} to simultaneously fit the spectrum shape and flux level. The
2M1207 B spectrum represents an example of how measurements of young, low mass object
atmospheres can advance the understanding of the atmospheric physics.

\subsection{Tucana-Horolgium systems}

An additional two substellar companions -- AB Pic B and GSC 080647 B -- are members of the
older $\sim$30 Myr Tucana-Horologium Association. AB Pic B is a wide orbit 5$\farcs$5
($\sim$260 AU) common proper motion companion \citep{Chauvin:2005c} to a K2 primary
\citep{Perryman:1997}. Based on a comparison of the system photometry with evolutionary
models \citep{Burrows:1997, Baraffe:2002} at 30 Myr, the mass of the companion is estimated
to be 13-14\Mjup\ \citep{Chauvin:2005c}, placing AB Pic B at the demarcation of planetary
masses and brown dwarf masses. The J,H,K SINFONI spectra have been used to estimate a
spectral type for the companion of L0-L1 and to compare each band of the spectrum with
theoretical atmosphere models \citep{Bonnefoy:2010}. In this paper, a model comparison of
the combined flux-calibrated J-K spectrum is presented, and the combined spectrum shows
differences with the models not apparent in the single band comparisons. Another
Tucana-Horologium pair with very similar spectral types to AB Pic is GSC 08047, which is
composed of a K3 primary with a confirmed substellar companion $\sim$3$\farcs$3 (279 AU)
from the host star \citep{Chauvin:2005a}. Previous spectra in the H- and K-bands taken with
VLT/ISAAC and VLT/NACO were compared with late-M objects to estimate a spectral type of M8
$\pm$ 2 \citep{Neuhauser:2004} and M9.5 $\pm$ 1 \citep{Chauvin:2005a}, slightly earlier than
the AP Pic companion. The new SINFONI spectra include the J-band and the IFU observations
are not subject to wavelength-dependent effects of differences in the slit centering for
target and standard that affected some of the previous spectra.

\subsection{Young Isolated Field Object}

As part of the large scale search and characterization program for L-dwarfs with 2MASS, an
isolated brown dwarf with indications of youth was identified with spectroscopy spanning 0.6
- 2.5 $\mu$m \citep{Kirkpatrick:2006}. In the infrared, there is a J-K SpeX spectrum with
similar resolution (R$\sim$1200) to the SINFONI data, and a NIRSPEC  J-band spectrum with
higher (R$\sim$2500 ) resolution, making 2M0141 an ideal case to test for variability such
as has been reported for TWA 30 \citep{Looper:2010}. For 2M0141, the spectral features of
weak alkali lines such as K I in the J-band and the triangular shape of the H-band caused by
a decrease in H$_2$ collision induced absorption provided evidence of a low surface gravity
object \citep{Kirkpatrick:2006}. The effective temperature and surface gravity were derived
from fits to atmospheric models generated with the PHOENIX code \citep{Kirkpatrick:2006},
and \teff\ and log(g) were compared with evolutionary models \citep{Baraffe:2002} to infer
an age range of 1-50 Myr and object mass of 6-25 Myr \citep{Kirkpatrick:2006}. Although
2M0141 currently has a less certain age, future measurements of the parallax and space
motion may determine membership in the Tucana-Horologium Association or the $\beta$ Pic
moving group. In this paper, the SINFONI spectra are compared with a larger set of
atmosphere models and with the earlier SpeX spectrum to search for variability.

\section{SINFONI Observations}

The observations were taken with the SINFONI (Spectrograph for INtegral Field Observations
in the Near Infrared) instrument \citep{Bonnet:2004,Eisenhauer:2003,Thatte:1998}, an
AO-equipped integral field spectrograph mounted at the Cassegrain focus of the VLT (UT4).
The target low mass companions are separated by less than a few arcseconds from the
primaries, making the primaries ideal sources for the AO correction. The observing sequence for all
targets included a set of exposures on the target with offset positions on the array. For
some targets, separate blank sky exposures were also recorded. Standard calibration
observations of an early-type or solar-type star followed each target sequence to correct
for telluric features and the instrument response. Additional calibrations were obtained to
measure the wavelength scale, the detector dark current, distortion, and quantum efficiency
variations. The data for this project were drawn from the ESO archive, and a summary of the
instrument configuration, observation details and project ID codes for each target is given
in Table \ref{table:obs}. In some cases, only the subset of higher quality data were used;
Table \ref{table:obs} lists only the observations used in this analysis.

There are four gratings available in SINFONI -- J, H, K, and H+K -- which produce spectral
resolutions of approximately 2000, 3000, 4000, and 1500, depending on the scale. For all
targets except GQ Lup B, a combination of J and H+K observations were taken. For GQ Lup B,
separate J, H, and K spectra were recorded. Among the three available spaxel (spatial pixel)
scales of 25mas, 100mas, and 250mas, the 100mas scale was employed for the CT Cha B and
2M1207A+B observations. The 100mas scale provides a field-of-view of 3$\farcs$0 $\times$
3$\farcs$0, and both the brown dwarf primary and planetary mass secondary for 2M1207 were
included within each observation. For the remaining target, the spaxel scale of 25mas was
used and the field-of-view at this scale is 0$\farcs$8 $\times$ 0$\farcs$8 which excluded
the primary host star from the observations. For one target, GQ Lup B, the diffraction spike
from the primary intersected the position of the target companion in a subset of the
observations.

\section{Data Analysis}

\subsection{ SINFONI data reduction}

The initial steps of the  data reduction for all targets were performed with the Gasgano
\citep{Gebbinck:2007} implementation of the SINFONI data reduction pipeline
\citep{Modigliani:2009,Dumas:2007}.  The processing began with the raw data and
calibration files in the ESO archive and is independent of previously reported results
summarised in Section 2. This approach was taken to ensure that all data used the same
software version and were reduced in a consistent way. Corrections for the dark current,
linearity response, and bad pixels were measured and applied to all data files. The
instrument optical distortion was measured with a set of fibers illuminating positions across
the field and the non-linear wavelength dispersion was measured with observations of an arc
lamp; both the spatial and spectral calibrations were applied to the data with Gasgano
routines. The calibrated two-dimensional data for each observation was converted into a
three-dimensional data cube consisting of an image at each wavelength slice.  Across the
entire wavelength range, a stack of $\sim$700-2000 wavelength slices images were recorded,
however the faintest target 2M1207B required binning in the J-band to produce a spectrum
with a higher signal-to-noise ratio.

After corrections for the detector and optics, the sky background emission was subtracted
from the offset position covering either blank sky or including the target at a different
position on the array in the ABBA pattern of observations. The individual sky-subtracted
cubes were aligned based on centroiding the target in the top wavelength slice and then
combined. To ensure the final spectrum traced a straight path through the data cube, each
wavelength slice of the final data cube was shifted and aligned, and then the final spectrum
was extracted. For the one target with both components on the array, 2M1207, it was
necessary to subtract the halo of the primary before extracting the spectrum of the
companion. Further details of the subtraction have been reported previously
\citep{Patience:2010}.

  \begin{figure}
   \centering
   \resizebox{\hsize}{!}{\includegraphics{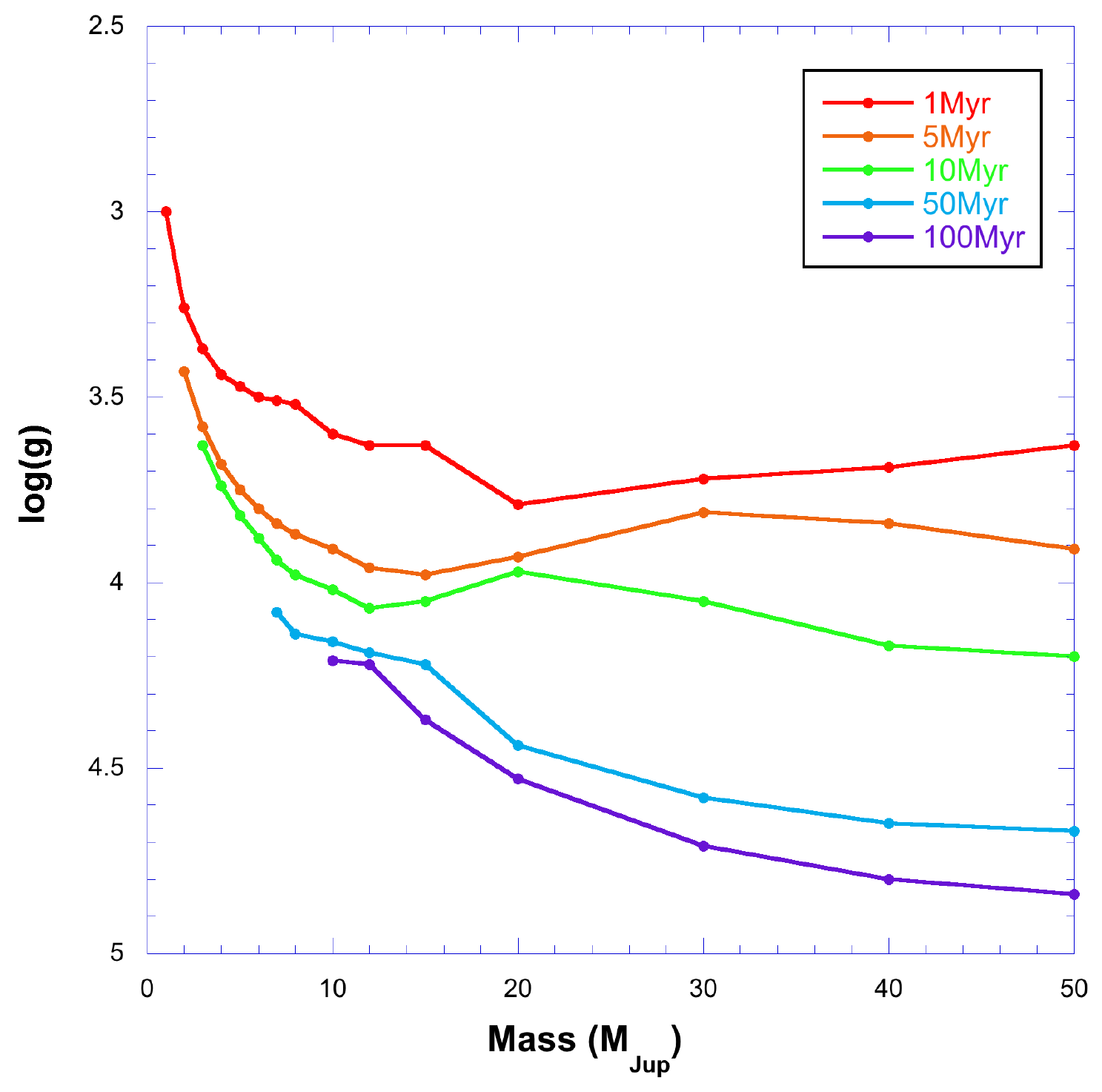}}
      \caption{The evolution in surface gravity over the 1-100 Myr age
      range for objects with masses of $\sim$ 1-50\Mjup. The
      values are taken from the DUSTY00 models \citep{Chabrier:2000}
      and show the systematic increase in log(g) over time as the
      objects contract. These tracks are used to determine the restricted range of 
      surface gravity to allow in model fits to our young targets.}
         \label{EvModels}
   \end{figure}
   
The standard stars observed after the targets were used to correct for telluric absorption
and the wavelength response of the detector by dividing each target spectrum by its
associated standard after interpolating the standard spectrum over intrinsic features. The
shape of the standard star continuum was removed by multiplication of a black body with the
same effective temperature as the standard. To flux calibrate the spectra, it was necessary
to rely on photometric measurements presented in Table \ref{table:phot} which were not
contemporaneous with the spectra. The observed spectra and the spectrum of Vega
\citep{Mountain:1985, Hayes:1985} were convolved with the 2MASS response curves
\citep{Carpenter:2001} to calculate the scaling factor required to match the reported
magnitude in a given filter. For the H+K spectra, the H-band photometry was used for flux
calibration, because the SINFONI data covers the full 2MASS H-band filter bandpass. For one
target, GQ Lup B, there is no reported H-band magnitude and the H and K spectra were
observed separately. In this case, the H-band magnitude was estimated by assuming the same
H-K color as DH Tau B, since GQ Lup B and DH Tau B both have a very similar J-K color and
similar young ages. Another target, CT Cha B, also had no H-band measurement, and, in this
case, the K-band magnitude was used to flux calibrate the H+K spectrum.

  \begin{figure}
   \centering
   \resizebox{\hsize}{!}{\includegraphics{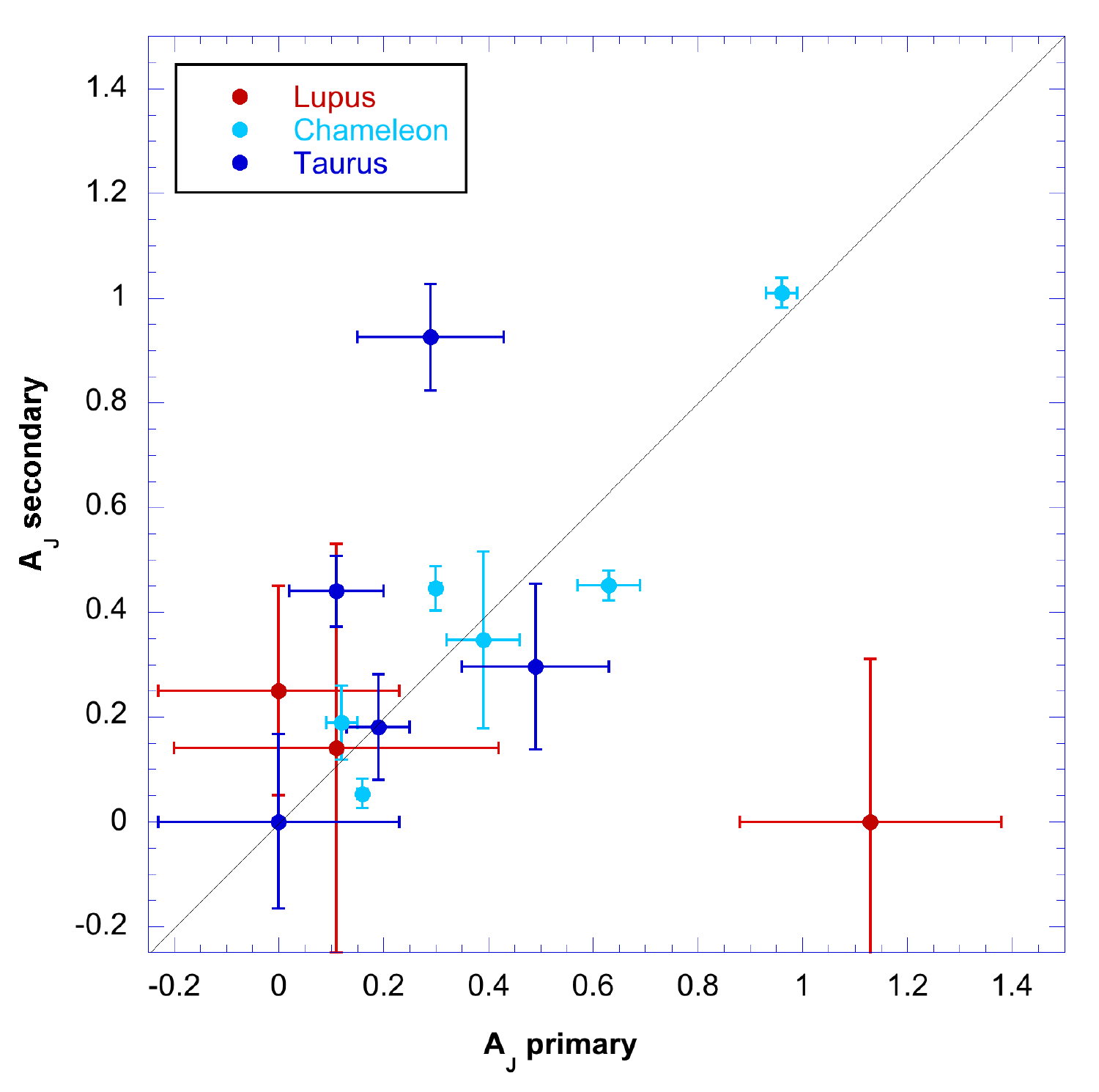}}
      \caption{The extinction A$_{J}$ of the secondary as a function
        of the A$_{J}$ of the primary for spatially resolved binary
        systems in Lupus (red), Chameleon (light blue), and Taurus
        (dark blue) based on observations of the individual components
      of each system \citep{Brandner:1997, White:2001,Prato:2003}. 
      These binaries have comparable
      separations to the target systems and are located in the same
      star-forming regions. The data show a trend of comparable
      extinction for each component. }
         \label{BinaryAj}
   \end{figure}

\begin{figure*}
\centering
\includegraphics[width=6cm]{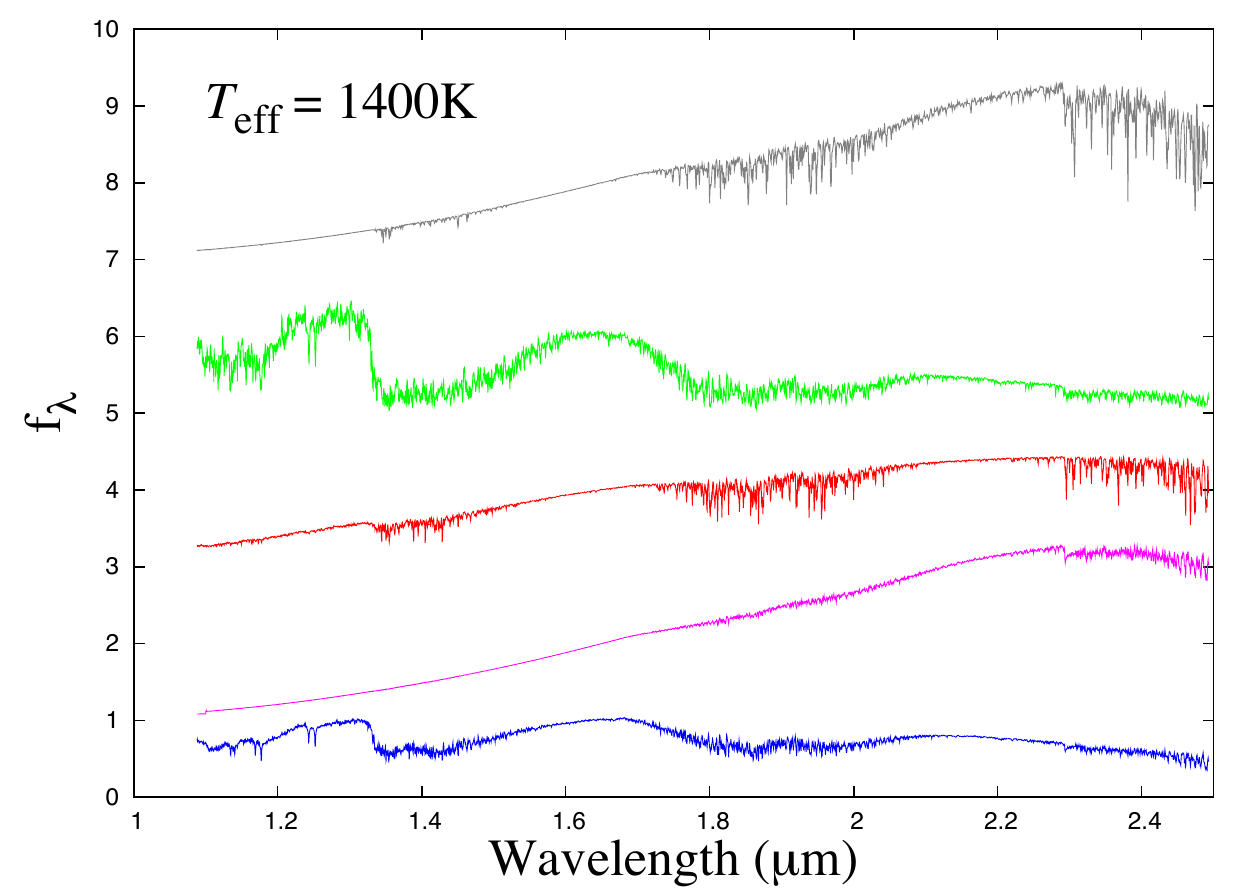}
\includegraphics[width=6cm]{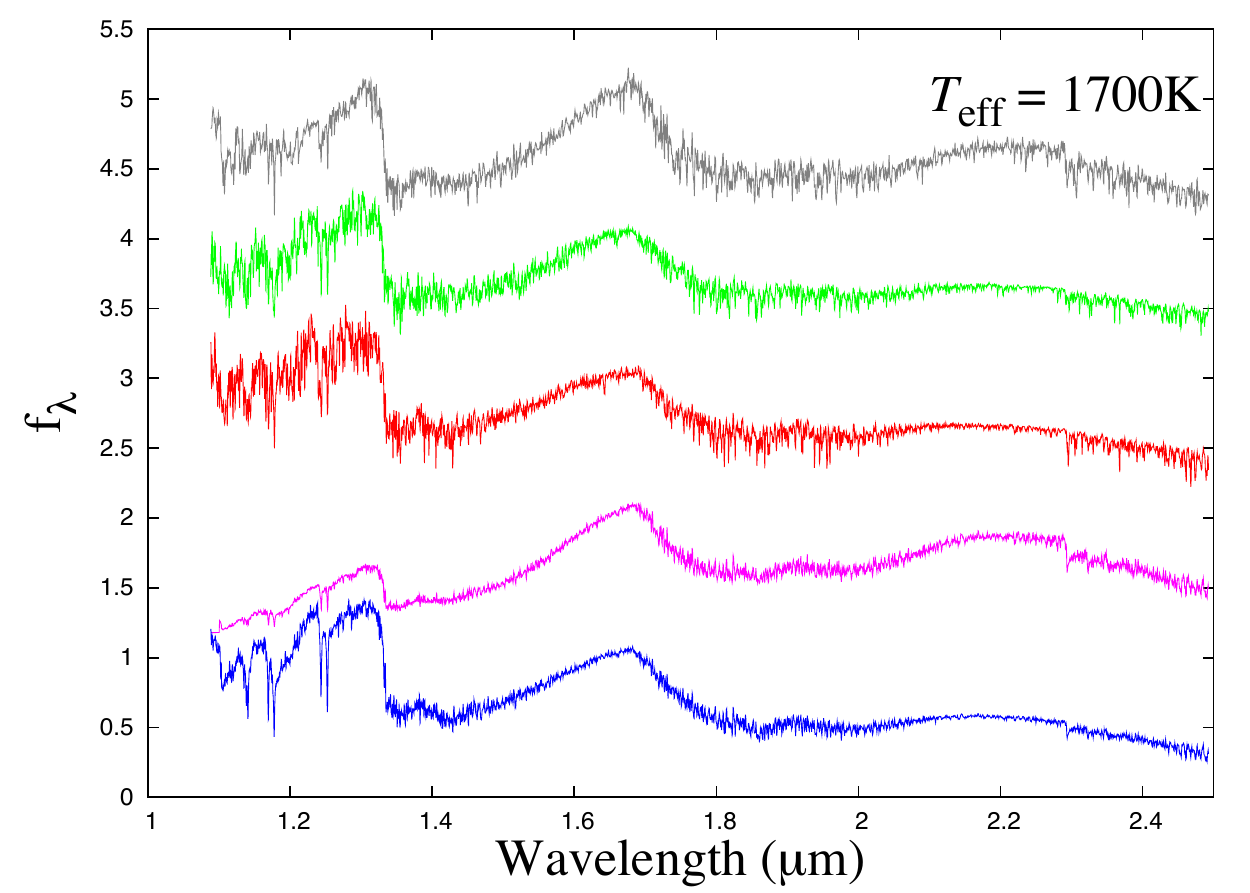}
\includegraphics[width=6cm]{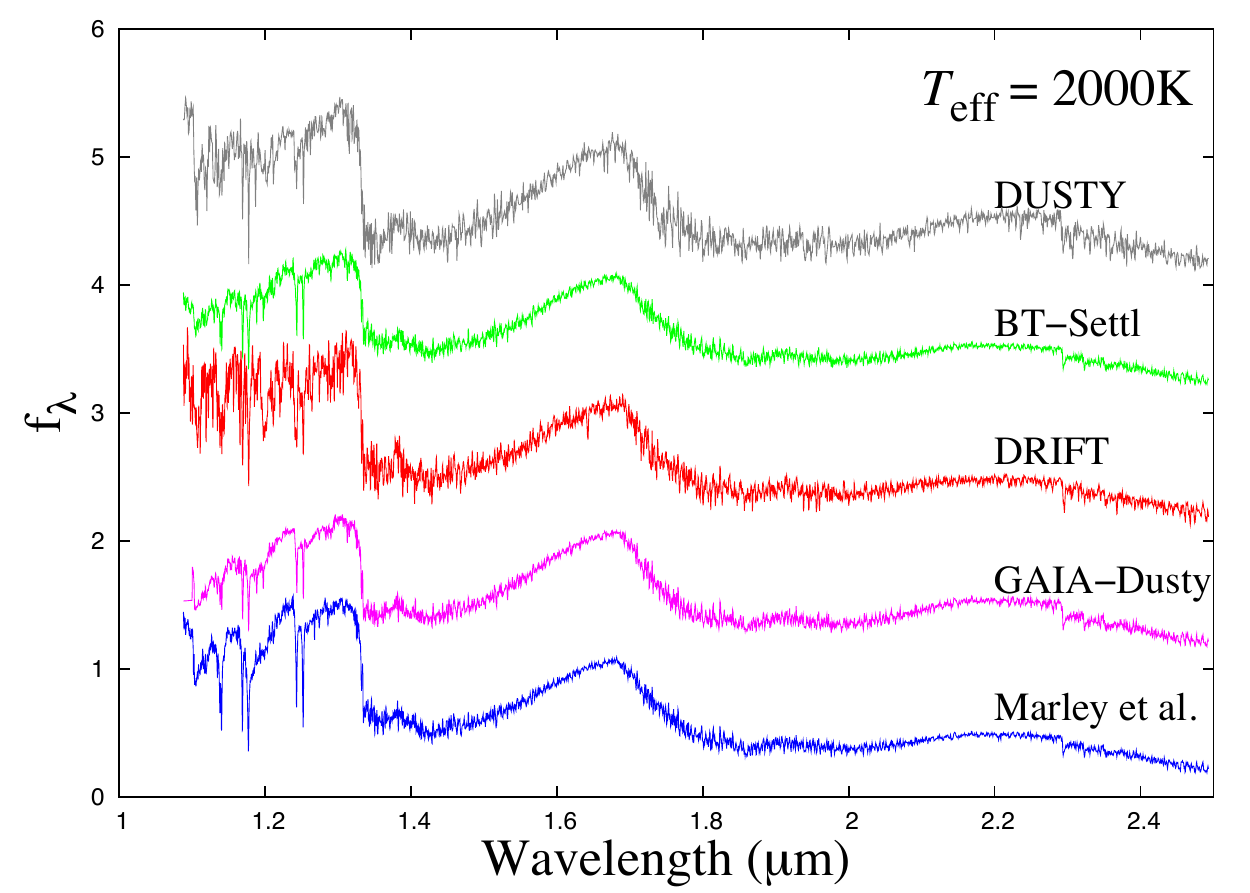}
\includegraphics[width=6cm]{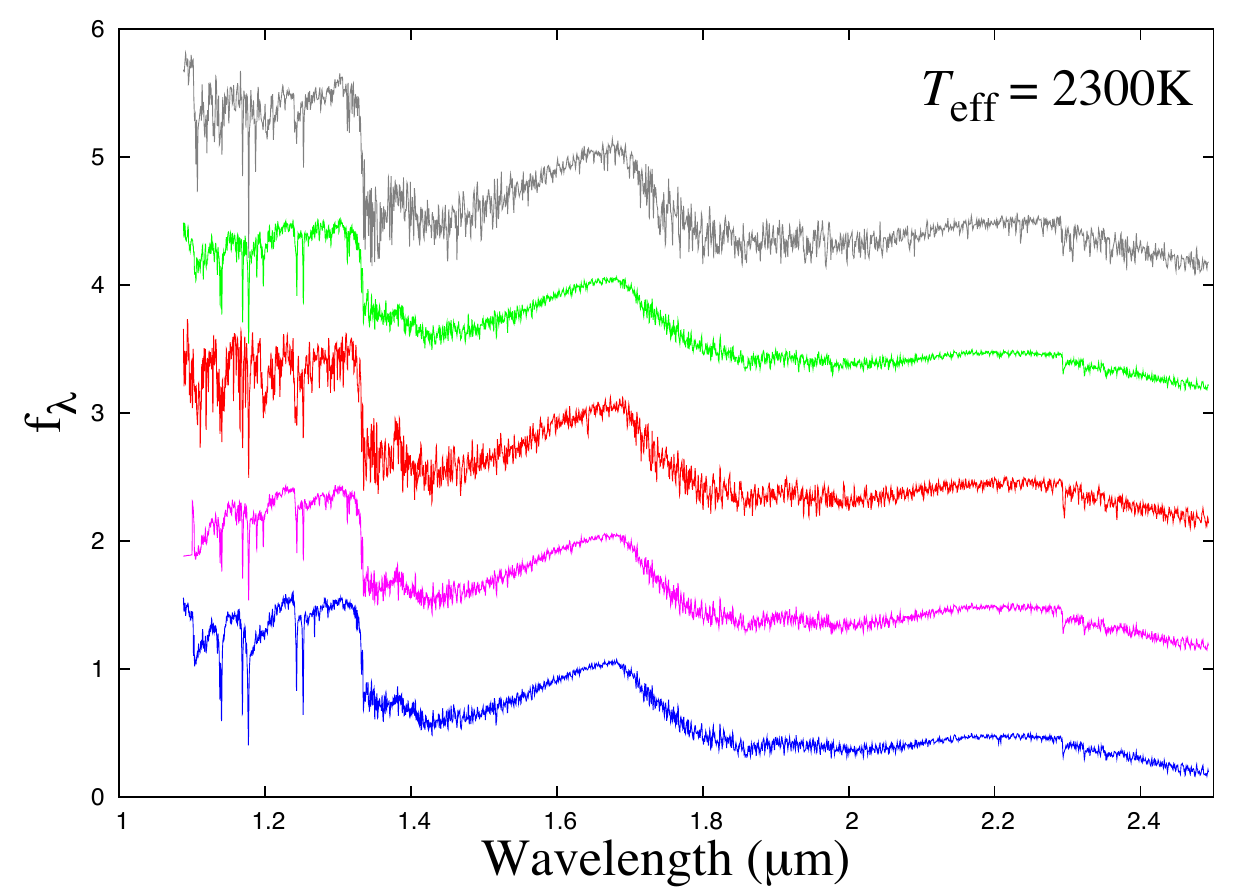}
\includegraphics[width=6cm]{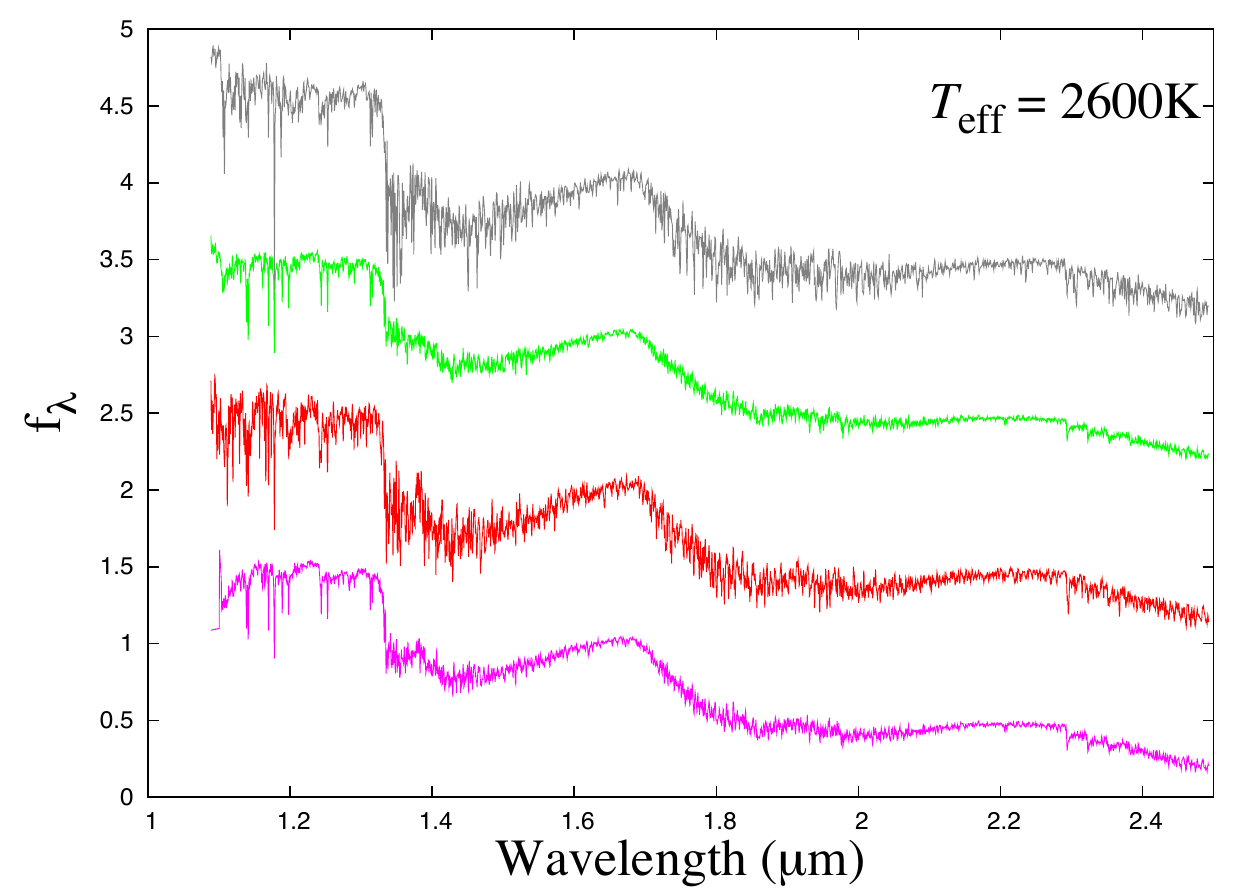}
\includegraphics[width=6cm]{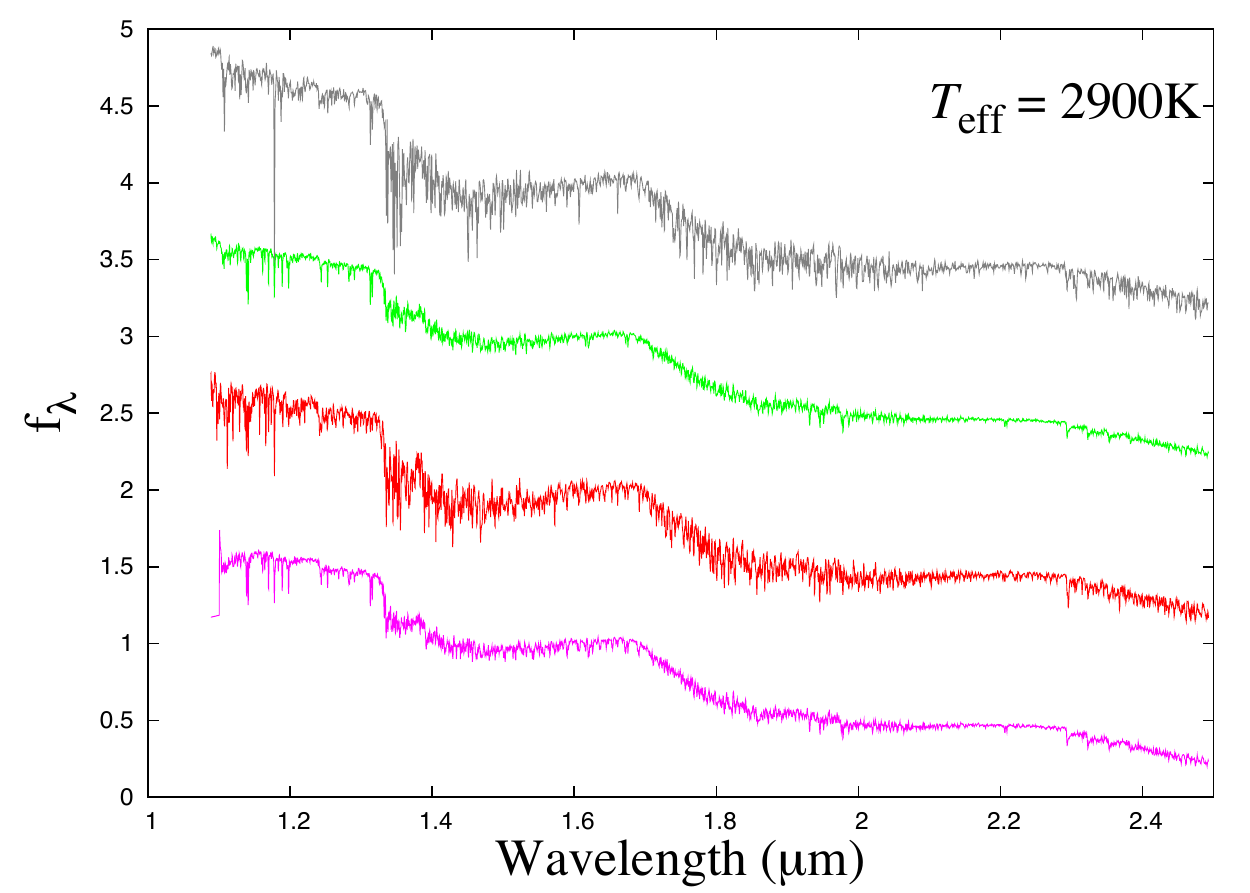}
\caption{For a fixed value of log(g)=4.5, the solar metallicity model 
spectra at temperatures ranging from 1400K (upper left) to 2900K (lower right) in
steps of 300K, corresponding to the target effective
temperatures. From top to bottom in each panel, the models are: DUSTY (grey),
BT-Settl (green), Drift-PHOENIX (red), Gaia-Dusty (purple), and Marley et al. 
(blue). The Marley et al. models do not include temperatures above
2400K, and the plots show models with f$_{sed}$=1, the dustiest
  set of models within this model grid.}
\label{ModelSpectra}
\end{figure*}

\begin{table*}
\caption{Cloud Models$^a$}             
\label{table:clouds}      
\renewcommand{\footnoterule}{}  
\centering          
\begin{tabular}{l c c c c c }     
\hline\hline     
\tiny
Physics  &  DUSTY  & BT-Settl & Drift-PHOENIX & Gaia-Dusty & Marley
et al.   \\ 
\hline      
Opacity source 		& 30 dust species & 43 dust species  & 7 dust species & updated DUSTY  		& 5 dust species    \\              
Gas-dust phase$^b$	&  $S=1$  	  &  $S=1.001$  &  $S=f(z,s)$ &  $S=1$  & $S=1$    \\       
Mixing    		&  none  	  & timescale from RHD  &
damping of turnover timescale & none  &  diffusive$^c$    \\       
 Settling/Rainout    	& none &  timescale comparison  &
 equation of motion & none &  parameterised in f$_{sed}$  \\       
Dust sizes 		&  $f(a)\propto a^{-3.5}$, (ISM)  & 
  timescale comparison  &  $f(a,z)$$^d$ &  $f(a)\propto
  a^{-3.5}$, (ISM)  &  log-normal $f(z,s)$$^e$ \\       
                        & 0.00625-0.24 $\mu$m &  0.00625-0.24 $\mu$m &
                         $10^{-6}-100$\,$\mu$m  &  0.00625-0.24
                        $\mu$m &  $10^{-5}-1$\,$\mu$m \\ 
Chemistry 		& equilibrium & equilibrium  & equilibrium$^f$
& equilibrium$^f$  & equilibrium except \\
   &   &   &  &  &  CO/CH$_4$, N$_2$/NH$_3$  \\            
\hline                  
\end{tabular}
\\  {\it Notes:} $^a$ PHOENIX \citep{Hauschildt:1999} DUSTY \citep{Allard:2001}, BT-Settl
\citep{Allard:2003, Allard:2010}, Drift-PHOENIX \citep{Helling:2008,Witte:2009}, 
Gaia-Dusty \citep{Rice:2010}, Marley et al. \citep{Ackerman:2001,
Marley:2003, Saumon:2008, Freedman:2008}  $^b$ $S$ quantifies the supersaturation
  ratio. $S=1$ for thermal equilibrium between the gas and dust. $^c$
  Parameterised in $k_{\rm{zz}}$.$^c$  $f(a,z)$ is the grain size
  distribution at height $z$ and provides the number of grains of size
  $a$. $^e$ $f(z,s)$ is the grain size distribution for a grain
  species $s$ at height $z$. $^f$ Updated chemistry treatment compared
  to DUSTY. 
\end{table*}
   
\subsection{Comparison with theoretical model grids}

The target spectra were compared with five sets of theoretical models, spanning a range of
effective temperature and surface gravity, as described in Section 5. The model spectra were
smoothed to the same resolution as the data and interpolated to the same dispersion before
performing a fitting procedure using the least-squares statistic defined in
\citet{Mohanty:2007} and allowing the model flux to be scaled to find the best fit. This
flux scaling is related to the object radius, enabling an estimate of the radius in addition
to the effective temperature and surface gravity. Since each wavelength step in the SINFONI
spectra is nearly constant over the JHK range, we do not employ the weighting technique of
\citet{Cushing:2008} which was developed for spectra covering a much larger range of
wavelengths.

Two sets of fits were performed to the model grids -- a fit including the full range of
log(g) values and a fit restricted to log(g)$\ge$3.0 and with an upper limit based on
evolutionary models \citep{Chabrier:2000} shown in Figure \ref{EvModels}. The targets
have known young ages, so the predicted evolution in surface gravity was used to set log(g)
limits: 4.0 for the companions to pre-Main Sequence stars, 4.5 for the TW Hydra members, and
5.0 for the 30-50 Myr targets.

A final factor affecting the slope of the observed spectrum is the amount of extinction to
each target. This is particularly important for the three objects in star-forming regions.
None of the secondaries have measured values of extinction, however, the primaries have
reported A$_V$ values. Rather than treat extinction as a free parameter, we use the results
of binary star component extinction comparisons to gauge the likelihood of very different
A$_J$ values for the primary and secondary. The secondary A$_J$ is shown as a function of
primary A$_J$ in Figure \ref{BinaryAj} for binary stars in the same regions as the targets
\citep{Brandner:1997, White:2001, Prato:2003}, and a clear trend of similar extinction
values is seen. Based on the binary stars, the targets in this sample are assumed to have
the same extinction as the primary, which is reported in Table \ref{table:sample}.

\section{Theoretical Models}

The data were compared with five grids of theoretical atmosphere models. Four of the models
represent different extensions of the PHOENIX \citep{Hauschildt:1999} code and incorporate a
range of dust properties and physical effects. In chronological order of development, the
PHOENIX-based models considered in this paper include: DUSTY \citep{Allard:2001}, BT-Settl
\citep{Allard:2003, Allard:2010}, Drift-PHOENIX \citep{Helling:2008,Witte:2009}, and
Gaia-Dusty \citep{Rice:2010}, with the latter three representing ongoing modelling efforts.
The earlier generation DUSTY model is included for comparison, although the BT-Settl and
Gaia-Dusty models have superseded the DUSTY models with improved treatments of abundances
and line lists. The independent model grid developed by Marley et al. \citep{Ackerman:2001,
Marley:2003, Saumon:2008, Freedman:2008} is also compared with the target spectra.

A summary of the physics incorporated into the five models is given in Table
\ref{table:clouds}; details of the models are given in the references listed above and the
comparison of models with two benchmark test cases is reported in \citep{Helling:2008b}. The
treatment of some physical processes such as convection is identical for all models, while
the approach to other factors such as condensate settling and chemistry varies between the
models. Since dust  plays a large role in determining the temperature-pressure
structure of the atmospheres in the temperature regime appropriate for the targets, the
number of dust species and dust size distributions involved in the calculations are listed,
along with notes on the mixing and settling included in the models.

An example of the synthetic spectra produced by each model over the same range of \teff\ for
a fixed value of log(g) = 4.5 is given in Figure \ref{ModelSpectra}. This value of log(g) was
selected because it is within the range predicted by evolutionary models for the ages of the
targets. For the Marley et al. spectra, the value of f$_{sed}$ is 1 which represents the
dustiest atmospheres. At the coolest temperature shown, 1400 K, there is a significant
difference in the synthetic shape of the H-band spectrum and the overall slope for the
different models. These effects are reduced by 1700K, but still impact the peak of the
J-band. By a temperature of 2000 K, the slope across the J-band ranges from flat to
increasing.  At hotter temperatures, the line strengths differentiate the models more than
the overall shape, and the atomic and metal hydride lines will be investigated for this
sample in a future paper (King et al. 2011, in prep.).

These model differences are primarily caused by the different treatments of condensate
settling and the differences in dust opacity. Typically, a higher overall dust opacity
results in a relatively lowered J-band flux and weakens the water bands while smoothing
their edges. The H band shape is dominated by the two adjacent water bands and the collision
induced absorption (CIA) which both grow stronger with decreasing dust opacities as flux
emerges from the deeper, hotter atmosphere resulting in a bluer spectrum.

\begin{table}
 \centering
 \caption{Model Grid Parameters}
\label{table:grid}
\begin{tabular}{lrcccc}
 \hline
 \hline
Model 			& \multicolumn{2}{c}{\teff}  &
\multicolumn{2}{c}{log(g)} & $\Delta$$\lambda$		\\
  					& range  & step  & range & step			\\
 \hline
DUSTY	        & 500-4000	& 100  & 3.5-6.0	& 0.5
& 2 $\AA$ \\
 BT-Settl$^a$			& 400-3000	& 100  & 3.5-5.5
 & 0.5  & 0.2 $\AA$  \\
Drift-PHOENIX		& 1000-3000	& 100  & 3.0-5.5	& 0.5
& 2 $\AA$\\
Gaia-Dusty	& 1400-3500	& 50    & 3.0-6.5	& 0.1  & 0.5 $\AA$\\
Marley et al.$^b$	& 900-2400	& 100  & 4.0-5.0	& 0.25 & R$\sim$50,000 \\
 \hline
\end{tabular}
\\{\it Note: }$^a$ for \teff=2000-3000K, the log(g) range is 0.0-5.5
$^b$ the wavelength step $\Delta$$\lambda$ is a function of $\lambda$
\end{table}

\begin{figure}
\centering
\includegraphics[scale=0.5]{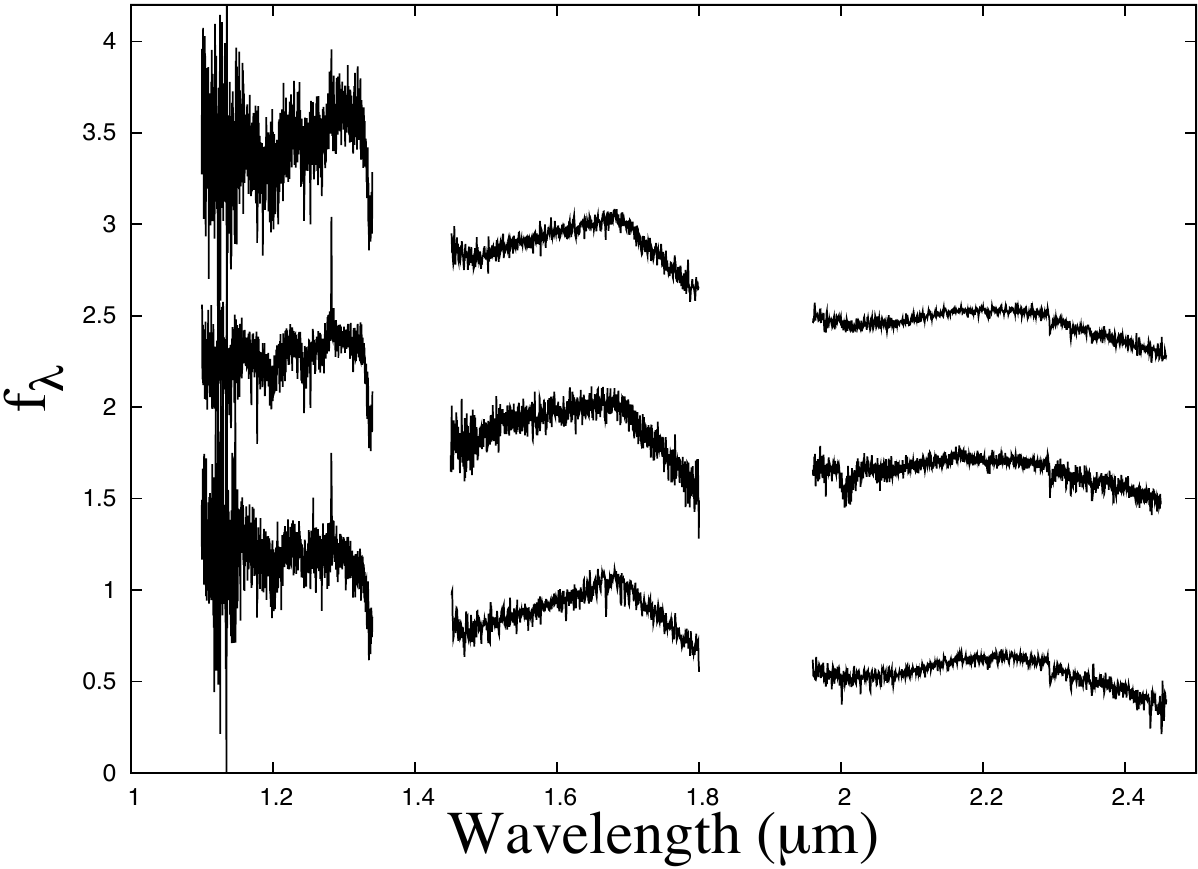}
\includegraphics[scale=0.5]{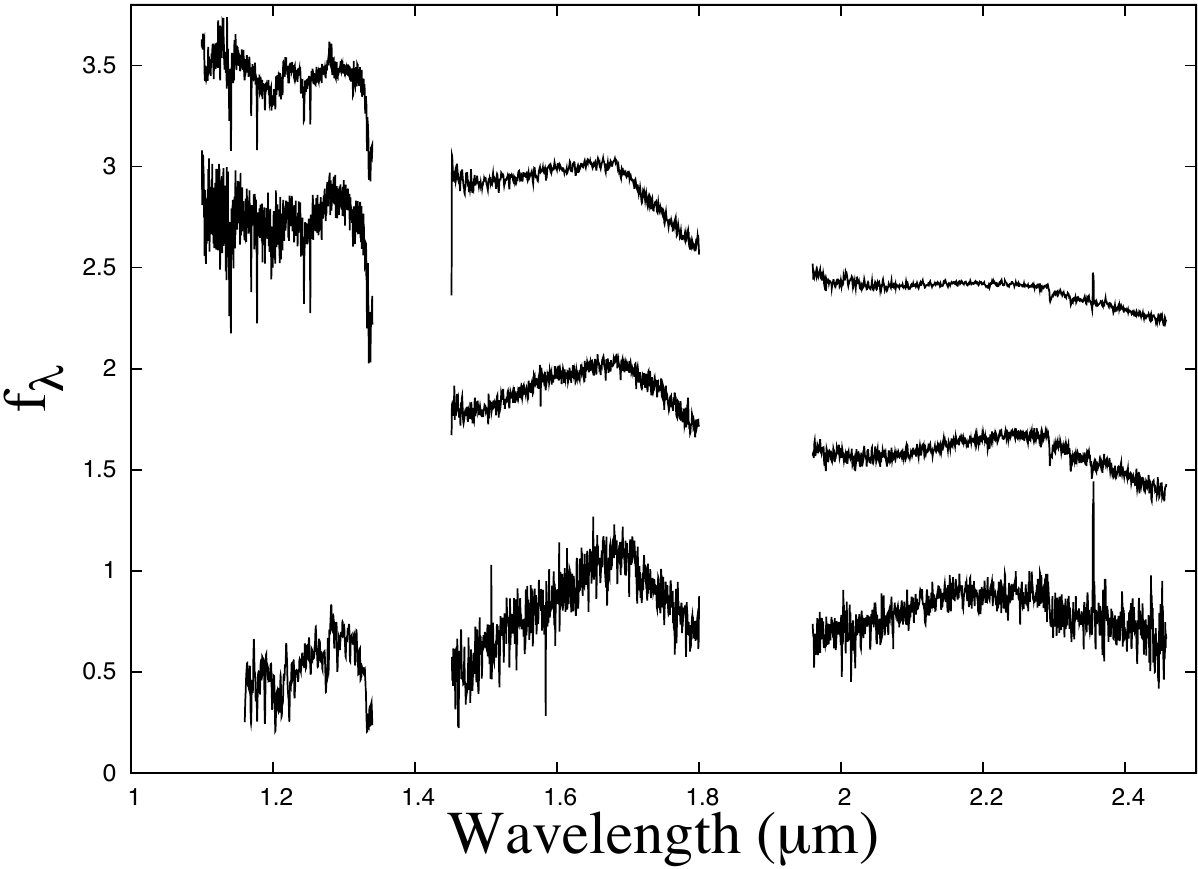}
\includegraphics[scale=0.5]{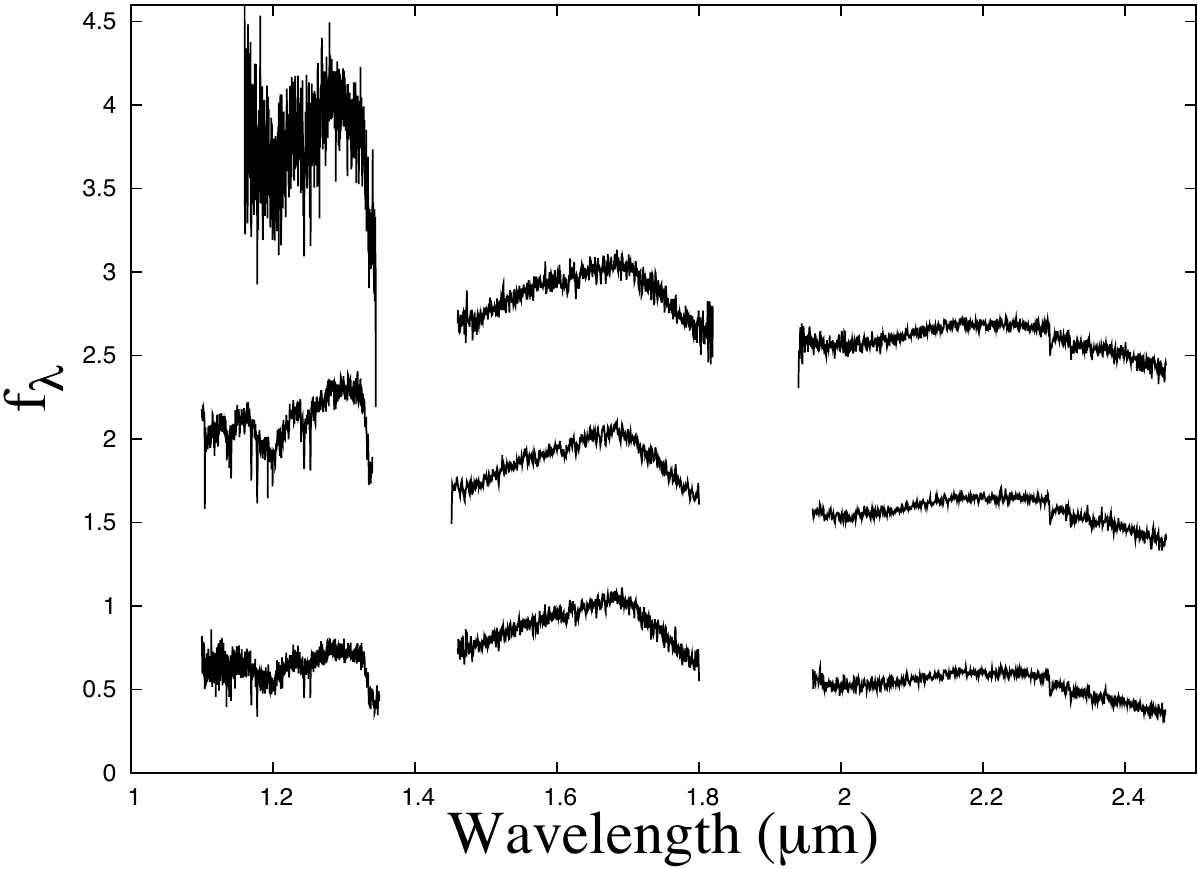}
\caption{SINFONI J,H,K spectra of the sample, separated into three age
groups: (top panel, top to bottom) the 1-3 Myr objects DH Tau B, GQ Lup B, and CT Cha
B; (middle panel. top to bottom) the 8 Myr objects 2M1207A, TWA 5B and 2M1207B;
(bottom panel, top to bottom) the 30-50 Myr objects GSC 08047B, 2M0141, and AB Pic
B. For each age group, the spectra are ordered from hottest (top) to
coolest (bottom), and an offset has been added to the flux for ease of
comparison. The spectra of the youngest subset have been dereddened based on the
extinction measured for the primaries.}
\label{ObsSpectra}
\end{figure}

Another difference between the models is apparent upon close comparison of the strong
absorption lines and bands between models calculated at different resolutions. In regions of
strong, varied absorption, models calculated at lower resolution (2 \AA) look `noisier',
i.e. a larger spread in the flux values over small wavelength
ranges (see Fig. \ref{ModelSpectra}).
For example, the DUSTY models show the largest `noise' of the spectrum over the full
temperature range, while the Marley et al., BT-Settl, and Gaia-Dusty models are typically
much narrower in the range of flux at a given wavelength.  These differences are a numerical
artifact induced when the convolution with a Gaussian instrument dispersion profile is
undersampled, resulting in lower fidelity model spectra for DUSTY and Drift-PHOENIX. 
The numerical origin of the difference was confirmed by calculating one grid (Gaia-Dusty)
with two initial resolutions and comparing the results.  Due to the
fact that the model fits in this paper
are based only on the spectral shape, the additional width in the DUSTY and Drift-PHOENIX
models does not impact this analysis, however the future comparison of spectral lines and
indices requires models with sufficient resolution to fit line strengths.

\begin{table*}
 \centering
 \caption{Best Fitting Atmopshere Model Parameters, Full Grid Comparison}
\label{table:fullgrid}
\begin{tabular}{lrlrlrlrlrll}
 \hline
 \hline
 Wavelength Range 				&
 \multicolumn{2}{c}{DUSTY} 	& \multicolumn{2}{c}{BT-Settl}
 & \multicolumn{2}{c}{Drift-PHOENIX}	& \multicolumn{2}{c}{Gaia-Dusty}
 & \multicolumn{3}{c}{Marley et al.}\\
  					& \teff\ (K) & \logg  &\teff\ (K) & \logg 	& \teff\ (K) & \logg & \teff\ (K) & \logg & \teff\ (K) & \logg&  $f_{\rm{sed}}$ \\
 \hline
\hline
\multicolumn{12}{c}{\bf{1-3 Myr Objects}} \\
\hline
\hline
  \multicolumn{12}{c}{DH Tau B$^a$} \\
 \hline
 JHK (1.10--2.46$\mu$m)		  & 2600	& 4.0  & 2400  & 2.5  & 2000	& 4.0& 2350  & 3.7  & $\geq$2400$^b$   & 4.0  & 2  \\
 J (1.10--1.34$\mu$m)			& 2400	& 4.0  & 2200  & 3.5  & 1600	& 3.5  & 2350  & 4.1  & $\geq$2400$^b$   & 4.0  & 2  \\
 HK (1.46--2.46$\mu$m)		& 2600	& 3.5  & 2500  & 3.5  & 1900	& 5.5  & 2550  & 3.8  & $\geq$2400$^b$   & 4.0  & 2  \\
 \hline
 \multicolumn{12}{c}{GQ Lup B$^a$} \\
 \hline
 JHK (1.10--2.46$\mu$m)		& 1900	& 6.0  & 1800  & 5.5  & 1700	& 3.5  & 2050  & 5.5  & 1600   & 4.5  & 1  \\
 J (1.10--1.34$\mu$m)			& 2500	& 4.0  & 2400  & 3.5  & 2600	& 4.0  & 2500  & 4.8  & $\geq$2400$^b$   & 4.0  & 2  \\
 HK (1.46--2.46$\mu$m)		& 1800	& 6.0  & 1700  & 4.5  & 1700	& 5.0  & 1800  & 5.5  & 1500   & 4.5  & 1  \\
 \hline
 \multicolumn{12}{c}{CT Cha B$^a$} \\
 \hline
 JHK (1.10--2.46$\mu$m)		& 1800	& 4.5  & 1600  & 4.0  & 1700	& 3.5  & 1950  & 4.1  & 1500   & 4.5  & 2  \\
 J (1.10--1.34$\mu$m)			& 1900	& 3.5  & 1400  & 5.0  & 1700	& 3.0  & 2000  & 3.9  & 1100   & 5.0  & 2  \\
 HK (1.46--2.46$\mu$m)		& 1900	& 6.0  & 2200  & 1.0  & 1800	& 5.0  & 1950  & 5.5  & 1700   & 4.5  & 1  \\
 \hline
\hline
\multicolumn{12}{c}{\bf{8 Myr Objects}} \\
\hline
\hline
\multicolumn{12}{c}{2MASS 1207 A} \\
 \hline
 JHK (1.10--2.46$\mu$m)		& 3100	& 5.5  & 2800  & 5.0  & $\geq$3000$^b$	& 5.0  & 2900  & 5.3  & $\geq$2400$^b$   & 5.0  & 1  \\
 J (1.10--1.34$\mu$m)			& 2900	& 5.0  & 2700  & 4.5  & $\geq$3000$^b$	& 5.0  & 2850  & 5.1  & $\geq$2400$^b$   & 4.0  & 2  \\
 HK (1.46--2.46$\mu$m)		& 3100	& 5.5  & 2800  & 5.5  & $\geq$3000$^b$	& 5.5  & 2950  & 5.5  & $\geq$2400$^b$   & 5.0  & 1  \\
 \hline
 \multicolumn{12}{c}{TWA 5 B} \\
 \hline
 JHK (1.10--2.46$\mu$m)		& 2500   & 3.5  & 2400  & 2.5  & 2200	& 3.0  & 2450  & 3.3  & $\geq$2400$^b$  & 4.0   & 2  \\
 J (1.10--1.34$\mu$m)			& 2700   & 4.5  & 2600  & 4.0  & 2800	& 4.5 & 2700  & 5.0  & $\geq$2400$^b$  & 4.0   & 2  \\
 HK (1.46--2.46$\mu$m)		& 1800   & 6.0  & 2200  & 1.0  & 1800	& 5.5  & 1900  & 5.5  & 1600  & 4.5   & 1  \\
 \hline
 \multicolumn{12}{c}{2MASS 1207 B} \\
 \hline
 JHK (1.10--2.46$\mu$m)		& 1600	& 3.5  & 1500  & 3.5  & 1500	& 5.0  & 1650  & 4.3  & 1100   & 5.0  & 1  \\
 J (1.10--1.34$\mu$m)			& 1700	& 4.5  & 2000  & 0.5  & 1700	& 4.0  & 1950  & 4.0  & 1500   & 4.0  & 1  \\
 HK (1.46--2.46$\mu$m)		& 1600	& 4.5  & 1500  & 3.5  & 1500	& 3.0  & 1650  & 4.7  & 1300   & 4.0  & 1  \\
 \hline
\hline
\multicolumn{12}{c}{\bf{30-50 Myr Objects}} \\
\hline
\hline
 \multicolumn{12}{c}{GSC 08047 B} \\
 \hline
 JHK (1.10--2.46$\mu$m)		& 2200	& 3.5  & 2200  & 1.0  & 1900	& 4.5  & 2250  & 3.1  & 2300   & 4.0  & 2  \\
 J (1.10--1.34$\mu$m)			& 2600	& 5.0  & 2200  & 3.5  & 1600	& 3.5  & 2300  & 4.2  & $\geq$2400$^b$   & 4.0  & 2  \\
 HK (1.46--2.46$\mu$m)		& 1800	& 6.0  & 1700  & 4.5  & 1700	& 4.0  & 1850  & 5.5  & 1600   & 4.5  & 1  \\
 \hline
 \multicolumn{12}{c}{2MASS 0141} \\
 \hline
 JHK (1.10--2.46$\mu$m)		& 1800	& 5.0  & 2000  & 3.0  & 1700	& 4.0  & 2000  & 5.5  & 1600   & 4.5  & 1  \\
 J (1.10--1.34$\mu$m)			& 2300	& 4.0  & 2200  & 3.5  & 1600	& 3.5  & 2300  & 4.9  & $\geq$2400$^b$   & 4.0  & 2  \\
 HK (1.46--2.46$\mu$m)		& 1800	& 6.0  & 2100  & 1.0  & 1800	& 5.0  & 1900  & 5.5  & 1600   & 4.5  & 1  \\
 \hline
\multicolumn{12}{c}{AB Pic B} \\
 \hline
 JHK (1.10--2.46$\mu$m)		& 1800   & 6.0  & 1600  & 3.5  & 1600	& 4.5  & 1850  & 5.5  & 1400  & 4.5   & 1  \\
 J (1.10--1.34$\mu$m)			& 2400   & 4.5  & 2300  & 4.0  & 1600	& 3.5  & 2400  & 4.9  & $\geq$2400$^b$  & 4.0   & 2  \\
 HK (1.46--2.46$\mu$m)		& 1900   & 6.0  & 2300  & 1.5  & 1800	& 5.0  & 1950  & 5.5  & 1700  & 5.0   & 1  \\
 \hline
\end{tabular}
\\  {\it Note: } $^a$ These star formation region members have been dereddened as described in Sect.\,4. 
$^b$ Note that this \teff\ falls at the edge of the grid used.  
\end{table*}

Table \ref{table:grid} shows the range of \teff\ and log(g) covered by the solar metallicity
models. The Marley et al. grids also incorporate a parameter f$_{sed}$ describing the
sedimentation efficiency, and the values considered for f$_{sed}$ in this grid are 1 and 2,
with the higher value associated with more efficient sedimentation from more rapid particle
growth and larger average particle sizes. In addition to the solar metallicity models
summarized in Table \ref{table:grid}, the Drift-PHOENIX grid also covered non-solar
metallicities with values of -0.6, -0.3, 0.0, and +0.3. The data are first compared with the
full grid summarized in Table \ref{table:grid} and then with a restricted subset with limits
on the log(g) values appropriate to the ages of the targets, based on evolutionary models
\citep{Chabrier:2000}. Finally, the impact of metallicity is explored with a comparison of
the data with the expanded grid with different metallicities for the Drift-PHOENIX model.

\section{Results and Discussion}

\subsection{SINFONI Spectra}

The J,H,K flux-calibrated SINFONI  spectra for the nine young substellar objects in the
sample are plotted in Figure \ref{ObsSpectra}, grouped in the three age bands and ordered in
decreasing temperature determined from the model comparisons described in Section 6.2.
Approximately half of the SINFONI data are previously unpublished: all wavelength bands of
DH Tau B, 2M1207B, and 2M0141 are newly reported along with the J- and H-bands of TWA 5B and
GSC 080647 B. For the youngest targets, the spectra have been dereddened according to the
reddening law of \citet{Cardelli:1989} using the extinction measured for the primary, as
described in Section 4.

\begin{table*}
 \centering
 \caption{Best Fitting Atmopshere Model Parameters, Restricted Grid Comparison}
\label{table:restgrid}
\begin{tabular}{lrlrlrlrlrll}
 \hline
 \hline
 Wavelength Range 			&
 \multicolumn{2}{c}{DUSTY} 	& \multicolumn{2}{c}{BT-Settl}
 & \multicolumn{2}{c}{Drift-PHOENIX}	& \multicolumn{2}{c}{Gaia-Dusty}
 & \multicolumn{3}{c}{Marley et al.}\\
  	& \teff\ (K) & \logg & \teff\ (K) & \logg  &\teff\
        (K) & \logg & \teff\ (K) & \logg    & \teff\ (K) & \logg &  $f_{\rm{sed}}$ \\
 \hline
\hline
\multicolumn{12}{c}{\bf{1-3 Myr Objects}} \\
\hline
\hline
 \multicolumn{12}{c}{DH Tau B$^a$} \\
 \hline
 JHK (1.10--2.46$\mu$m)		& 2600	& 4.0  & 2400  & 3.0  & 2000	& 4.0  & 2350  & 3.7  & $\geq$2400$^b$   & 4.0  & 2  \\
 J (1.10--1.34$\mu$m)			& 2400	& 4.0  & 2200  & 3.5  & 1600	& 3.5  & 2350  & 4.0  & $\geq$2400$^b$   & 4.0  & 2  \\
 HK (1.46--2.46$\mu$m)		& 2600	& 3.5  & 2500  & 3.5  & 2700	& 4.0  & 2550  & 3.8  & $\geq$2400$^b$   & 4.0  & 2  \\
 \hline
 \multicolumn{12}{c}{GQ Lup B$^a$} \\
 \hline
 JHK (1.10--2.46$\mu$m)		& 1900	& 3.5  & 1900  & 3.5  & 1700	& 3.5  & 2050  & 3.5  & 1500   & 4.0  & 1  \\
 J (1.10--1.34$\mu$m)			& 2500	& 4.0  & 2400  & 3.5  & 2600	& 4.0  & 2500  & 4.0  & $\geq$2400$^b$   & 4.0  & 2  \\
 HK (1.46--2.46$\mu$m)		& 1700	& 4.0  & 1700  & 3.5  & 1700	& 4.0  & 1750  & 4.0  & 1500   & 4.0  & 1  \\
 \hline
\multicolumn{12}{c}{CT Cha B$^a$} \\
 \hline
 JHK (1.10--2.46$\mu$m)		& 1900	& 4.0  & 1600  & 4.0  & 1700	& 3.5  & 1950  & 4.0  & 1500   & 4.0  & 1  \\
 J (1.10--1.34$\mu$m)			& 1900	& 3.5  & 1400  & 4.0  & 1700	& 3.0  & 2000  & 3.9  & 1000   & 4.0  & 1  \\
 HK (1.46--2.46$\mu$m)		& 2500	& 3.5  & 2300  & 3.0  & 1800	& 4.0  & 2350  & 3.1  & 2400   & 4.0  & 2  \\
 \hline
\hline
\multicolumn{12}{c}{\bf{8 Myr Objects}} \\
\hline
\hline
\multicolumn{12}{c}{2MASS 1207 A} \\
 \hline
 JHK (1.10--2.46$\mu$m)		& 3200	& 4.0  & 2800  & 4.5  & $\geq$3000$^b$	& 4.5  & 2900  & 4.5  & $\geq$2400$^b$   & 4.5  & 1  \\
 J (1.10--1.34$\mu$m)			& 2800	& 4.5  & 2700  & 4.5  & 2900	& 4.5  & 2800  & 4.5  & $\geq$2400$^b$   & 4.0  & 2  \\
 HK (1.46--2.46$\mu$m)		& 3100	& 4.5  & 3000  & 3.0  & $\geq$3000$^b$	& 4.5  & 3050  & 4.5  & $\geq$2400$^b$   & 4.5  & 1  \\
 \hline
 \multicolumn{12}{c}{TWA 5 B} \\
 \hline
 JHK (1.10--2.46$\mu$m)		& 2500   & 3.5  & 2300  & 3.0  & 2200	& 3.0  & 2450  & 3.3  & $\geq$2400$^b$  & 4.0   & 2  \\
 J (1.10--1.34$\mu$m)			& 2700   & 4.5  & 2600  & 4.0  & 2800	& 4.5  & 2700  & 4.5  & $\geq$2400$^b$  & 4.0   & 2  \\
 HK (1.46--2.46$\mu$m)		& 2500   & 3.5  & 2000  & 3.0  & 1700	& 4.5  & 2250  & 3.1  & 1600  & 4.5   & 1  \\
 \hline
 \multicolumn{12}{c}{2MASS 1207 B} \\
 \hline
 JHK (1.10--2.46$\mu$m)		& 1600	& 3.5  & 1500  & 3.5  & 1500	& 4.0  & 1650  & 4.3  & 1300   & 4.0  & 1  \\
 J (1.10--1.34$\mu$m)			& 1700	& 4.5  & 2000  & 3.0  & 1700	& 4.0  & 1950  & 4.0  & 1500   & 4.0  & 1  \\
 HK (1.46--2.46$\mu$m)		& 1600	& 4.5  & 1500  & 3.5  & 1500	& 3.0  & 1650  & 4.5  & 1300   & 4.0  & 1  \\
 \hline
\hline
\multicolumn{12}{c}{\bf{30-50 Myr Objects}} \\
\hline
\hline
 \multicolumn{12}{c}{GSC 08047 B} \\
 \hline
 JHK (1.10--2.46$\mu$m)		& 2200	& 3.5  & 2200  & 3.0  & 1900	& 4.5  & 2250  & 3.1  & 2300   & 4.0  & 2  \\
 J (1.10--1.34$\mu$m)			& 2600	& 5.0  & 2200  & 3.5  & 1600	& 3.5  & 2300  & 4.2  & $\geq$2400$^b$   & 4.0  & 2  \\
 HK (1.46--2.46$\mu$m)		& 1700	& 5.0  & 1700  & 4.5  & 1700	& 4.0  & 1750  & 5.0  & 1600   & 4.5  & 1  \\
 \hline
\multicolumn{12}{c}{2MASS 0141} \\
 \hline
 JHK (1.10--2.46$\mu$m)		& 1800	& 5.0  & 2000  & 3.0  & 1700	& 4.0  & 2000  & 5.0  & 1600   & 4.5  & 1  \\
 J (1.10--1.34$\mu$m)			& 2300	& 4.0  & 2200  & 3.5  & 1600	& 3.5  & 2300  & 4.9  & $\geq$2400$^b$   & 4.0  & 2  \\
 HK (1.46--2.46$\mu$m)		& 1800	& 5.0  & 2000  & 3.0  & 1800	& 5.0  & 1850  & 5.0  & 1600   & 4.5  & 1  \\
 \hline
 \multicolumn{12}{c}{AB Pic B} \\
 \hline
 JHK (1.10--2.46$\mu$m)		& 1700   & 5.0  & 1600  & 3.5  & 1600	& 4.5  & 1800  & 4.8  & 1400  & 4.5   & 1  \\
 J (1.10--1.34$\mu$m)			& 2400   & 4.5  & 2300  & 4.0  & 1600	& 3.5  & 2400  & 4.9  & $\geq$2400$^b$  & 4.0   & 2  \\
 HK (1.46--2.46$\mu$m)		& 2500   & 3.5  & 2300  & 3.0  & 1800	& 5.0  & 2300  & 5.0  & 1700  & 5.0   & 1  \\
 \hline
 
\end{tabular}
\\  {\it Note: } $^a$ These star formation region members have been dereddened as described in Sect.\,4. 
$^b$ Note that this \teff\ falls at the edge of the grid used.  
\end{table*}

The overall spectral shape is dominated by the H$_2$O absorption at both sides of each
wavelength band and the CO absorption band at the red edge of the K-band. All the object
spectra exhibit a distinct triangular shape in the H-band portion of the spectrum, a
signature of young substellar objects. As noted in other spectra of young objects, the peak
of the K-band also occurs at a longer wavelength near 2.3$\mu$m than is typical for older
field L-dwarfs \citep{Mohanty:2007}. Two of the coolest objects, 2M1207 B and AB Pic B, show
an unusually depressed J-band portion of the spectrum relative to the spectra of the other
substellar objects of the same age.

\subsection{Comparison with Atmosphere and Evolutionary Models}

\subsubsection{Full Model Grid}

Each object spectrum was compared with a grid of the five solar-metallicity models
summarized in Section 5 -- DUSTY, BT-Settl, Drift-PHOENIX, Gaia-Dusty, and Marley et al. 
The range in the parameters of \teff\ and log(g) are summarized in Table \ref{table:grid}.
The model fits were performed on each full J,H,K spectrum and the J and H+K subsets to
investigate the variation of inferred physical properties on the wavelength coverage of the
data. The results of the model fits for \teff\ and log(g) are given in Table
\ref{table:fullgrid}, divided by the age groups and ordered by average \teff\ from the J,H,K
fit.

Based on the J,H,K spectrum fits, the effective temperatures range from 1100 K to 3100 K and
log(g) of 1.0 to 6.0, considering the extremes of any given model fit. Some of the values
for log(g) are outside the bounds expected from evolutionary models, and the fits also do
not show a clear trend of increasing log(g) with increasing age. The lack of a clear
correlation between log(g) and age may simply be a consequence of the coarseness of the
grids and the dominant impact of temperature over surface gravity in shaping the overall
spectral energy distribution, as opposed to specific gravity-sensitive features. The range
of masses included in the sample also complicates the comparison of log(g) and age.

  \begin{figure*}
   \centering
\includegraphics[scale=0.75]{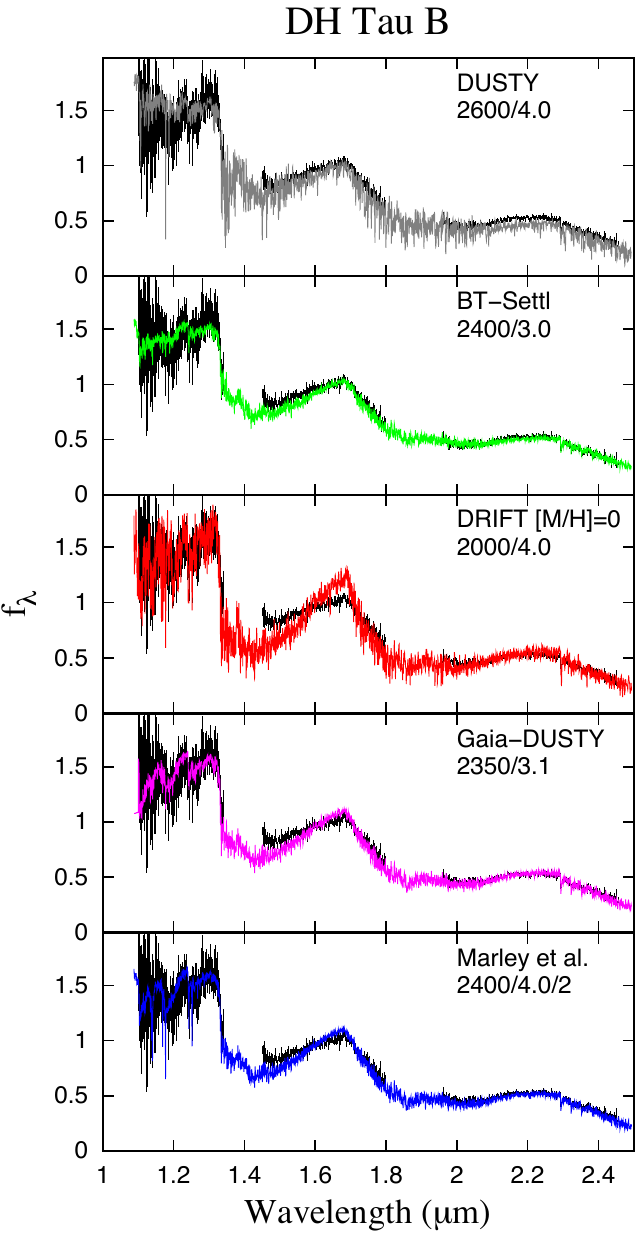}
\includegraphics[scale=0.75]{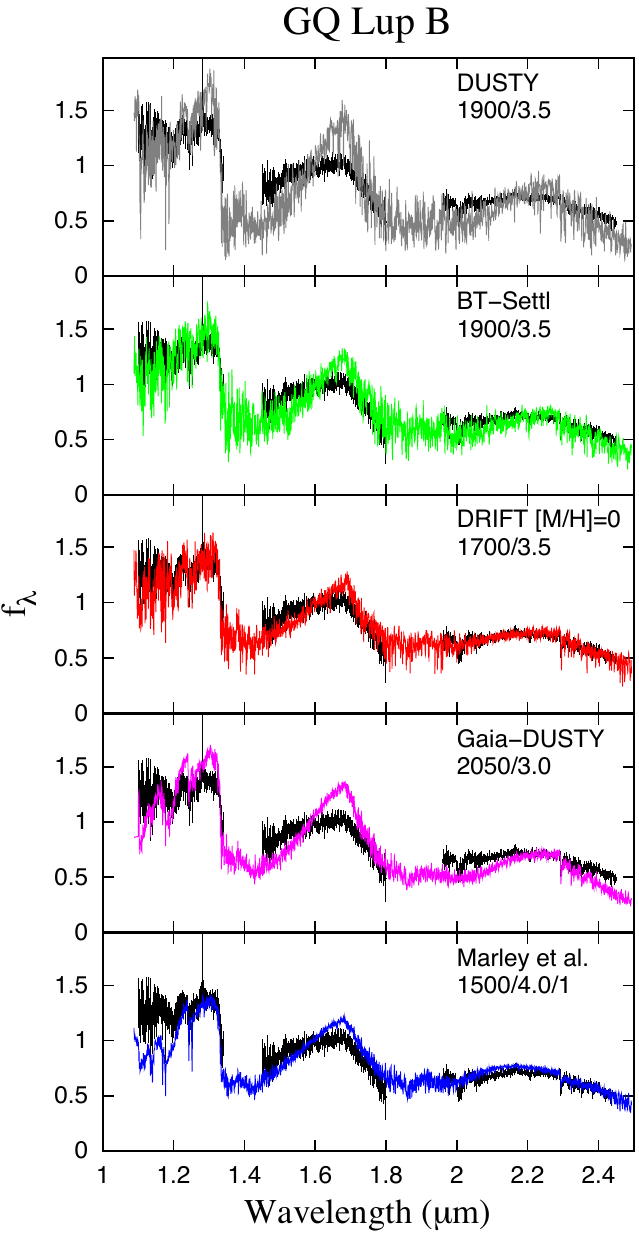}
\includegraphics[scale=0.75]{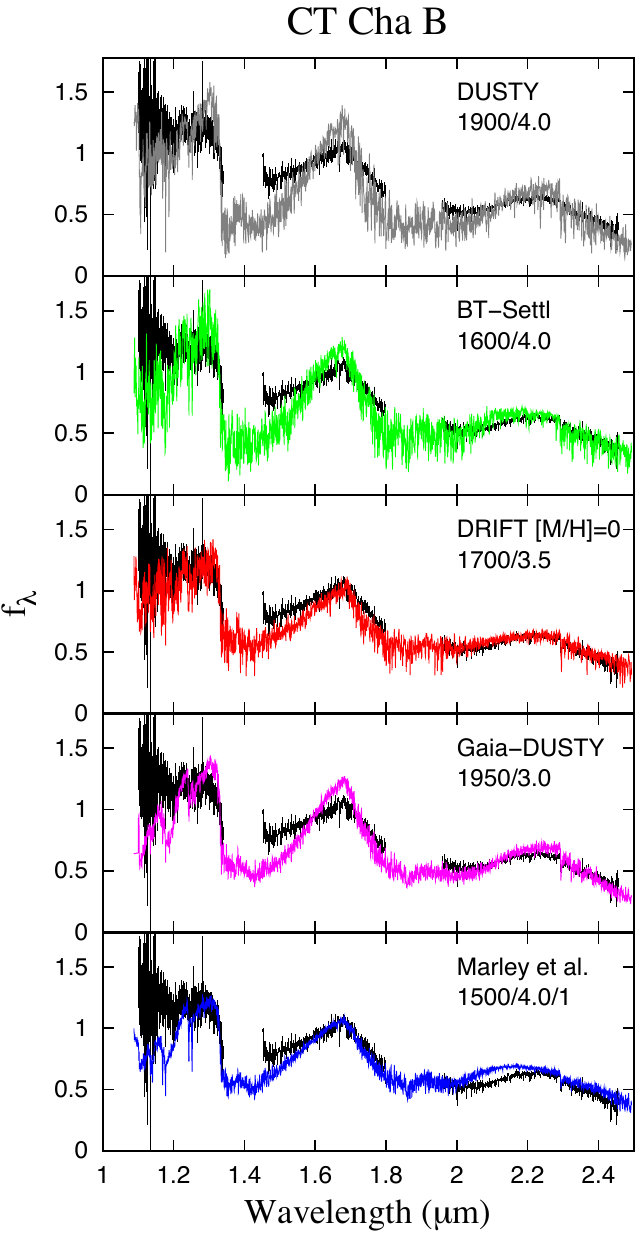}
      \caption{The SINFONI spectra ( black lines) and best-fit
          model (colored lines) to
      the full spectrum is plotted for the star-forming region
      targets: DH Tau B (left), GQ Lup B (middle), and CT Cha B
      (right). The restricted log(g) model grids are used.}
         \label{Grid1}
   \end{figure*}

  \begin{figure*}
   \centering
\includegraphics[scale=0.75]{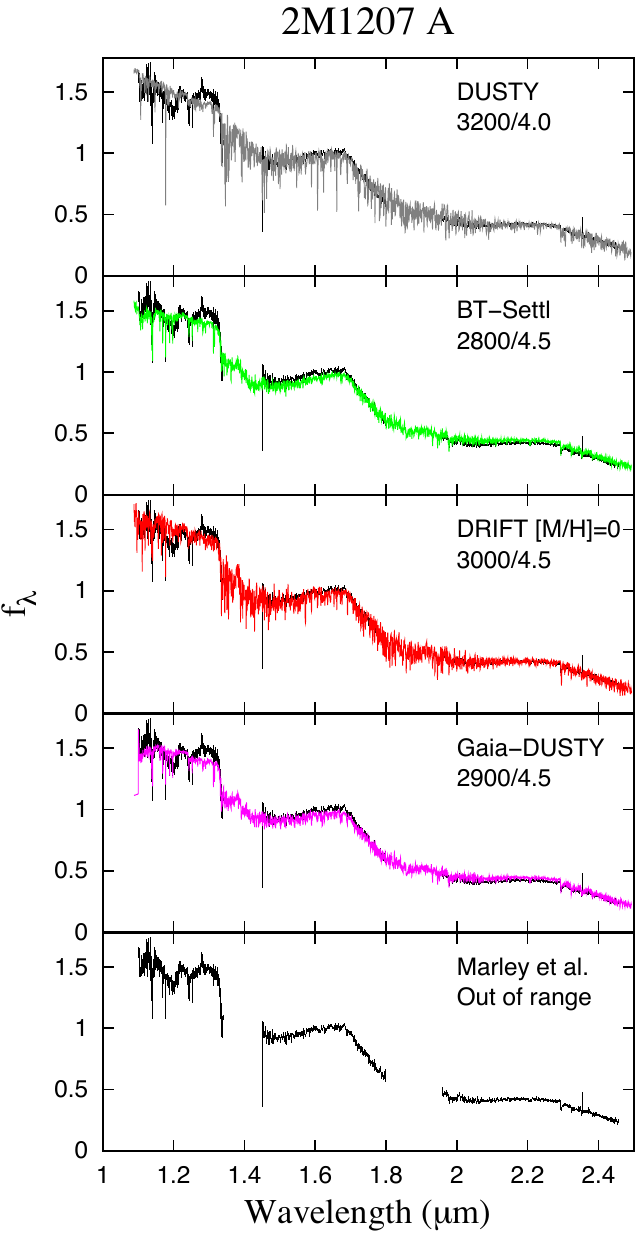}
\includegraphics[scale=0.75]{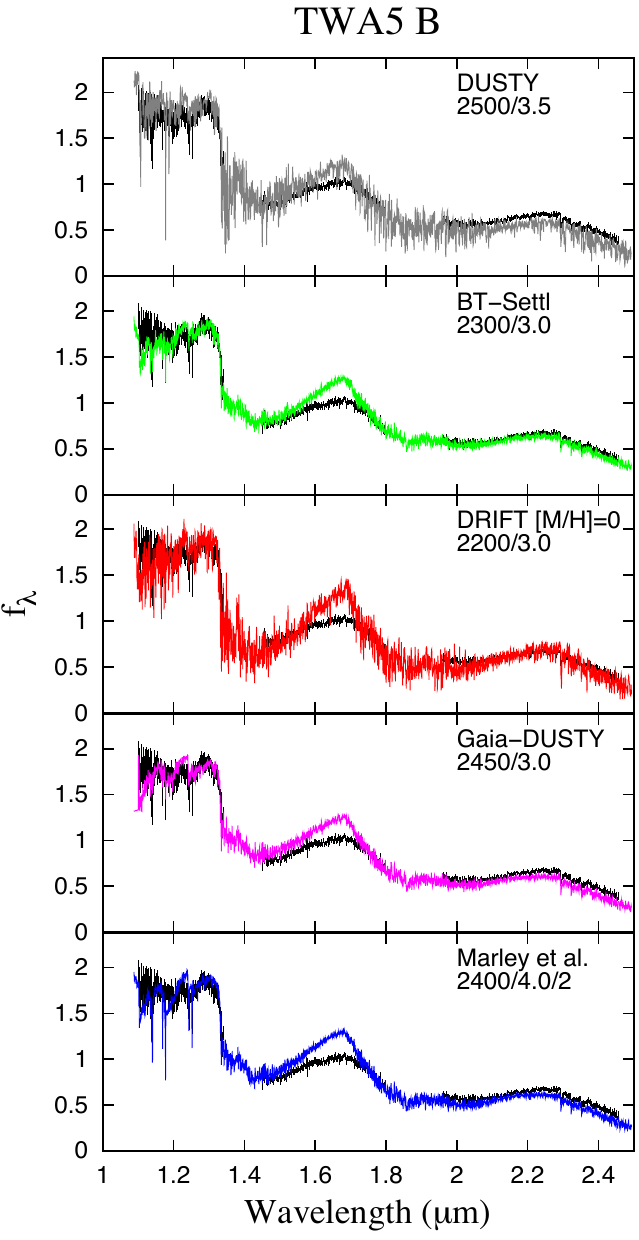}
\includegraphics[scale=0.75]{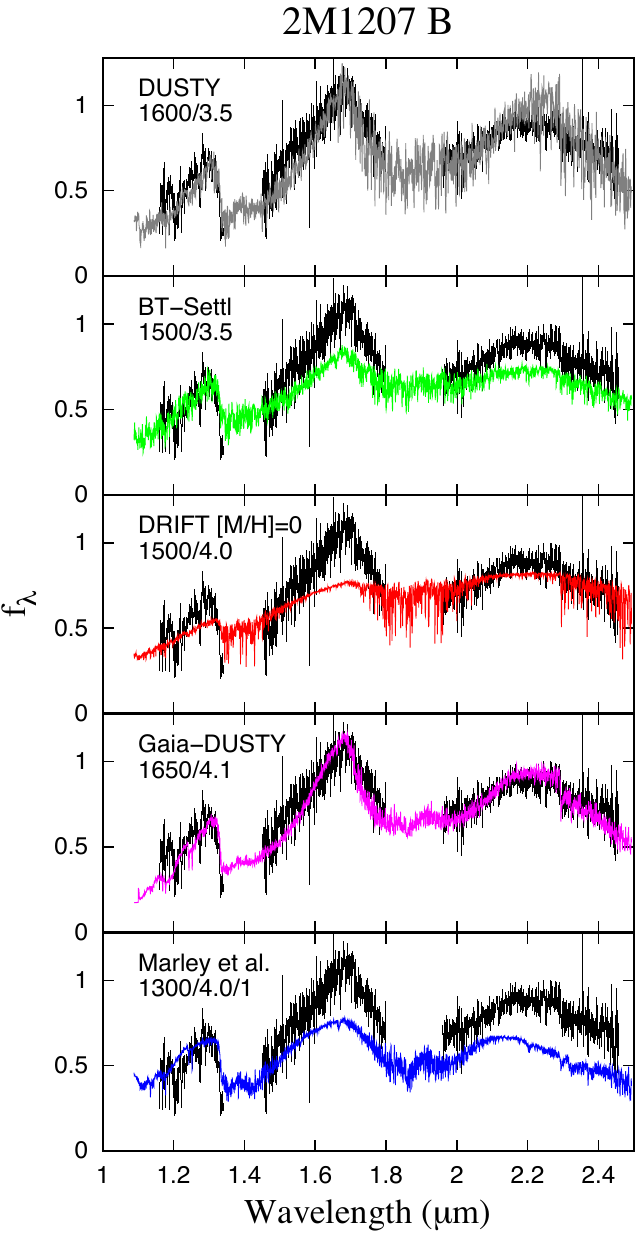}
      \caption{The SINFONI spectra (black lines) and best-fit model (colored lines) to
      the full spectrum is plotted for the TW Hydra member
      targets: 2M1207 A (left), TWA 5B (middle), and 2M1207 B
      (right). The restricted log(g) model grids are used.}
         \label{Grid2}
   \end{figure*}

  \begin{figure*}
   \centering
\includegraphics[scale=0.75]{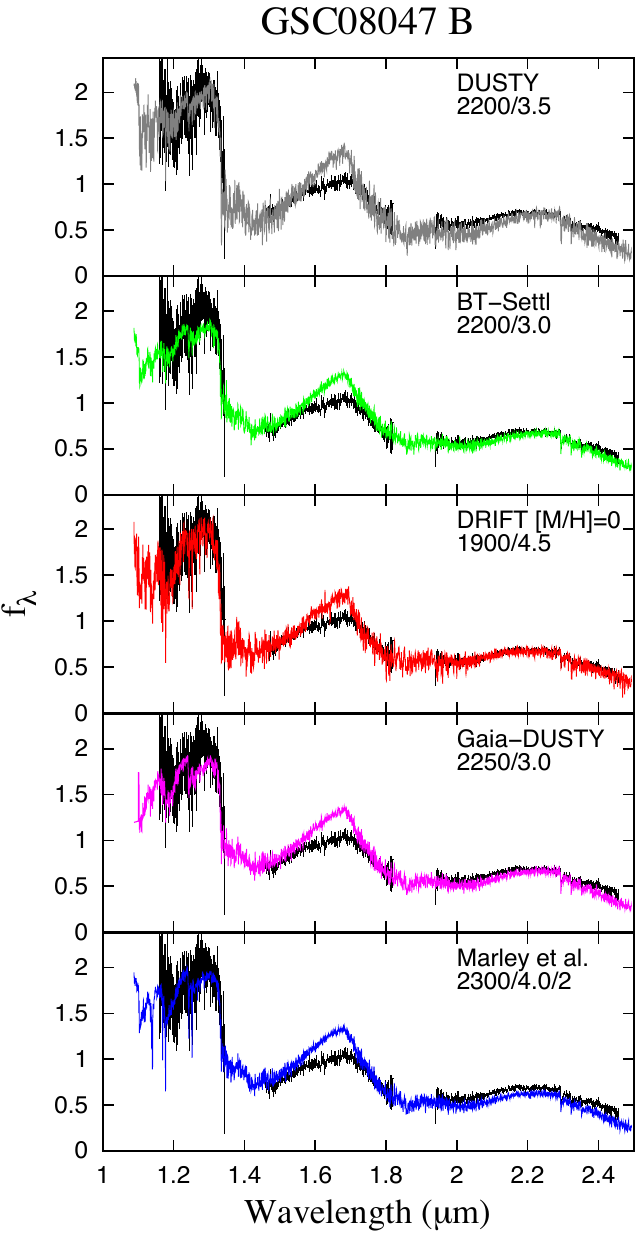}
\includegraphics[scale=0.75]{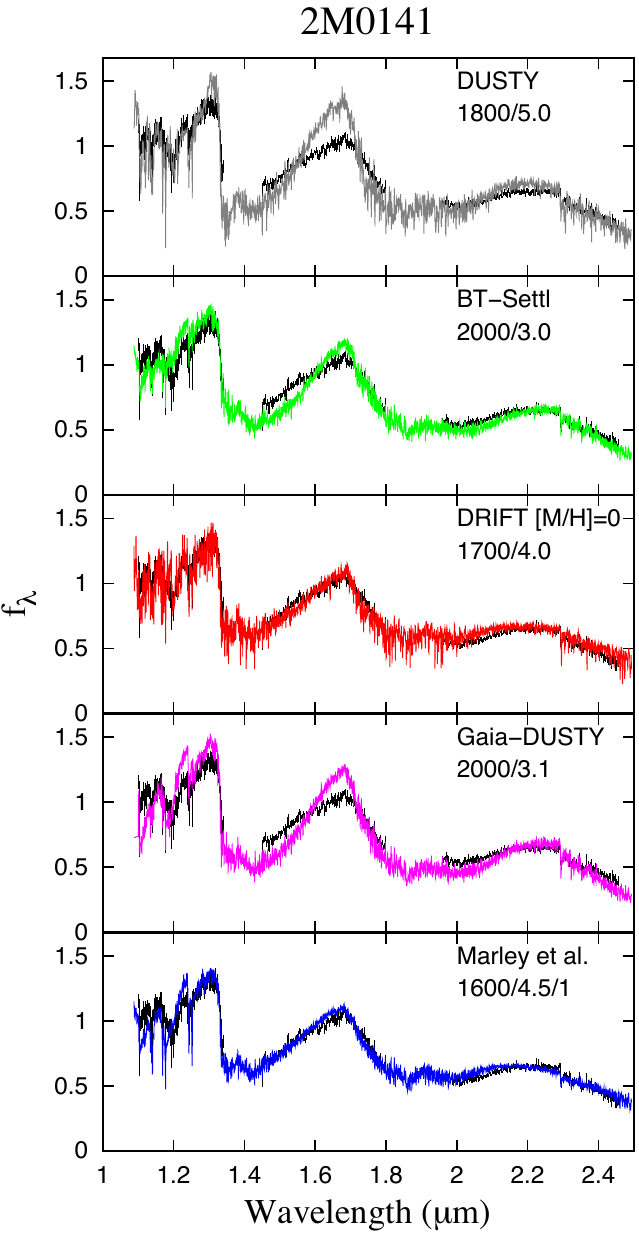}
\includegraphics[scale=0.75]{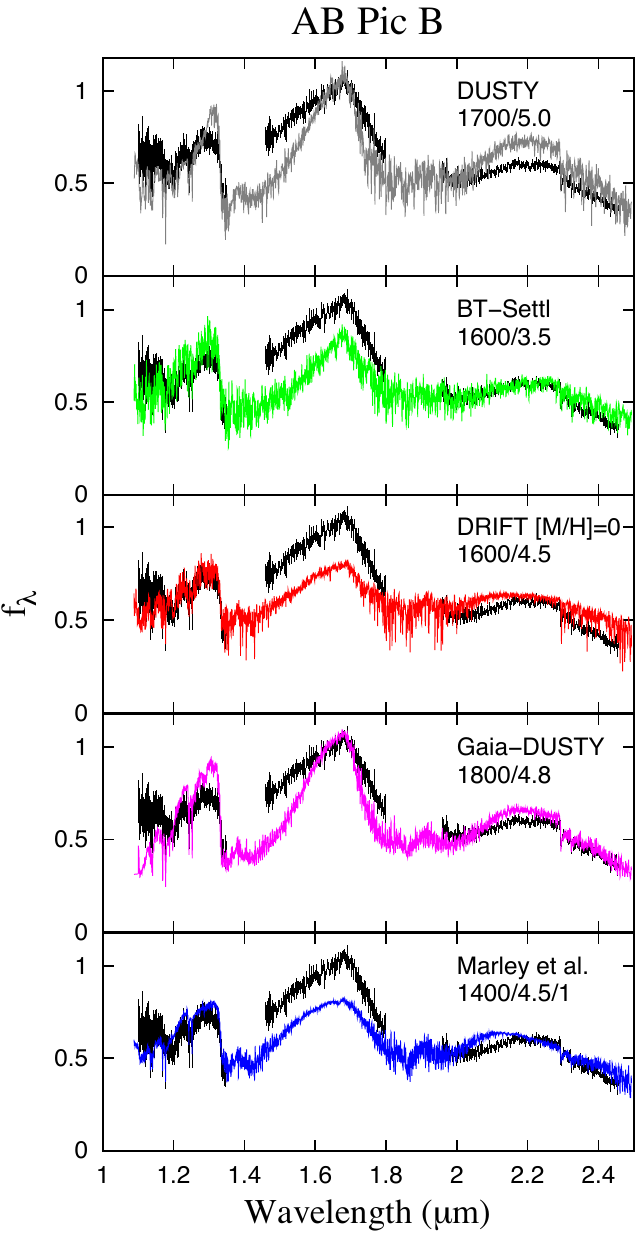}
      \caption{The SINFONI spectra (black lines) and best-fit model (colored lines) to
      the full spectrum is plotted for the Tucana-Horologium and Field 
      targets: GSC 08047 B (left), 2M0141 (middle), and AB Pic B
      (right). The restricted log(g) model grids are used.}
         \label{Grid3}
   \end{figure*}
   
  \begin{figure*}
   \centering
   \includegraphics[scale=0.38]{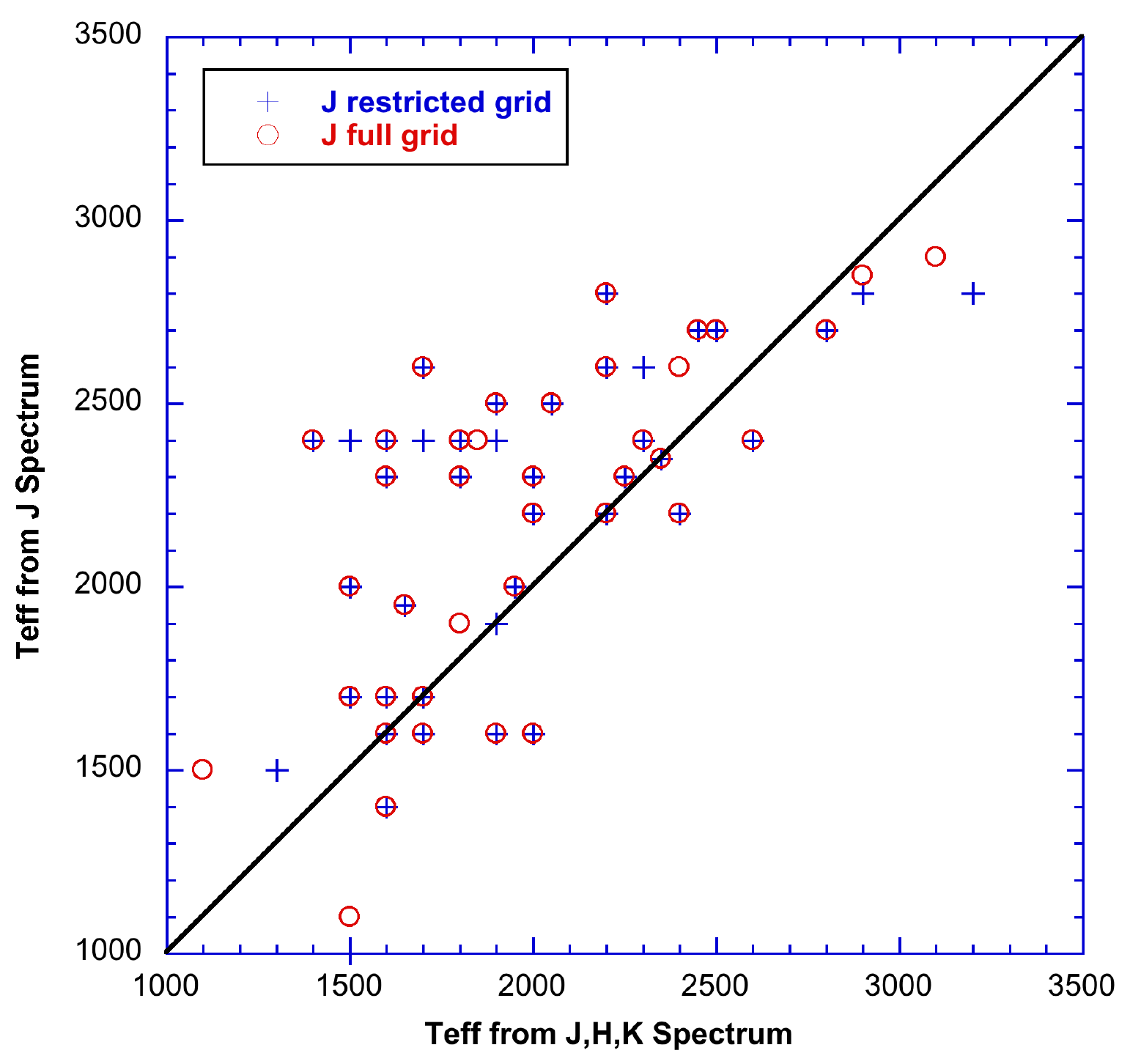}
   \includegraphics[scale=0.38]{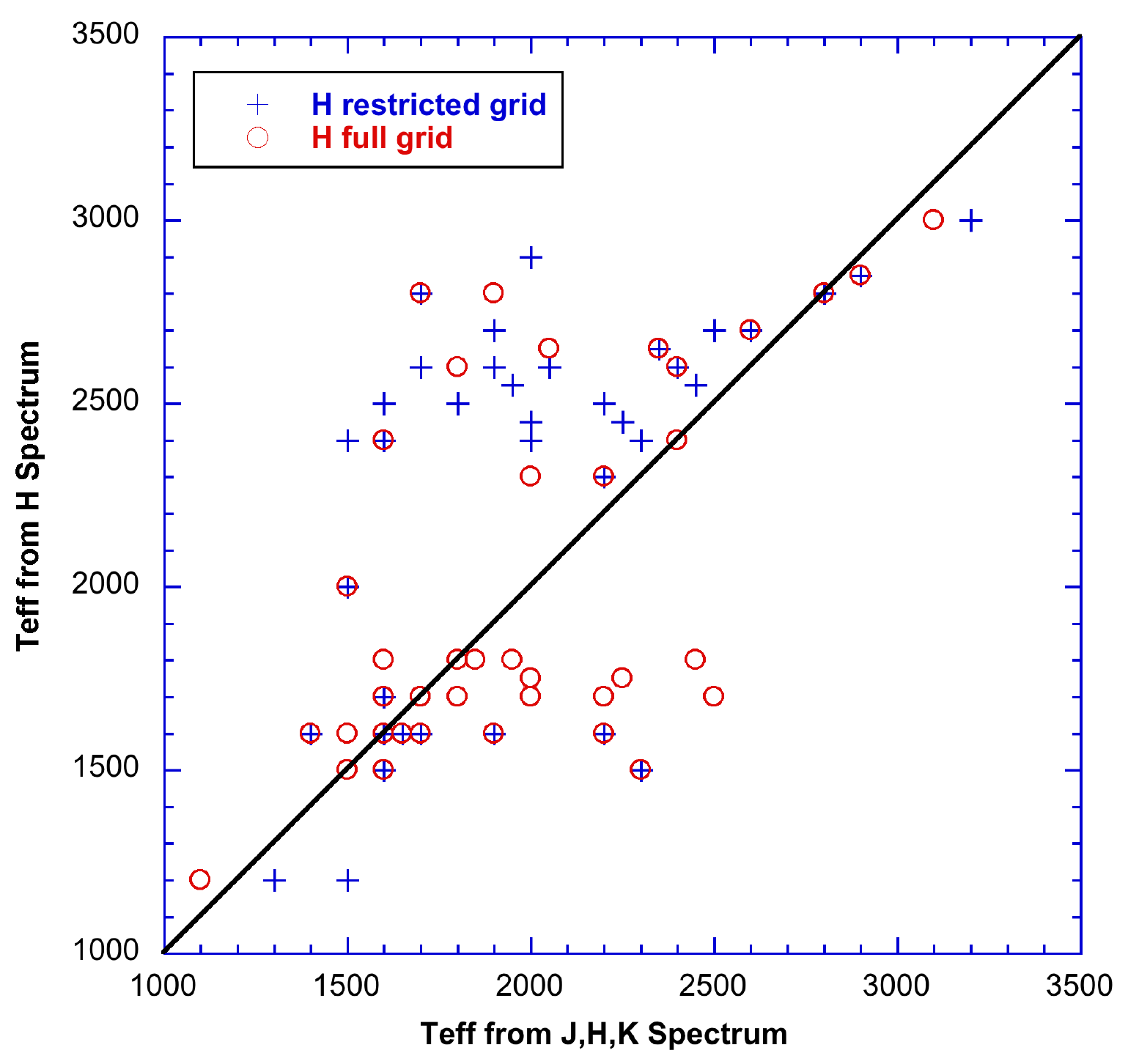}
   \includegraphics[scale=0.38]{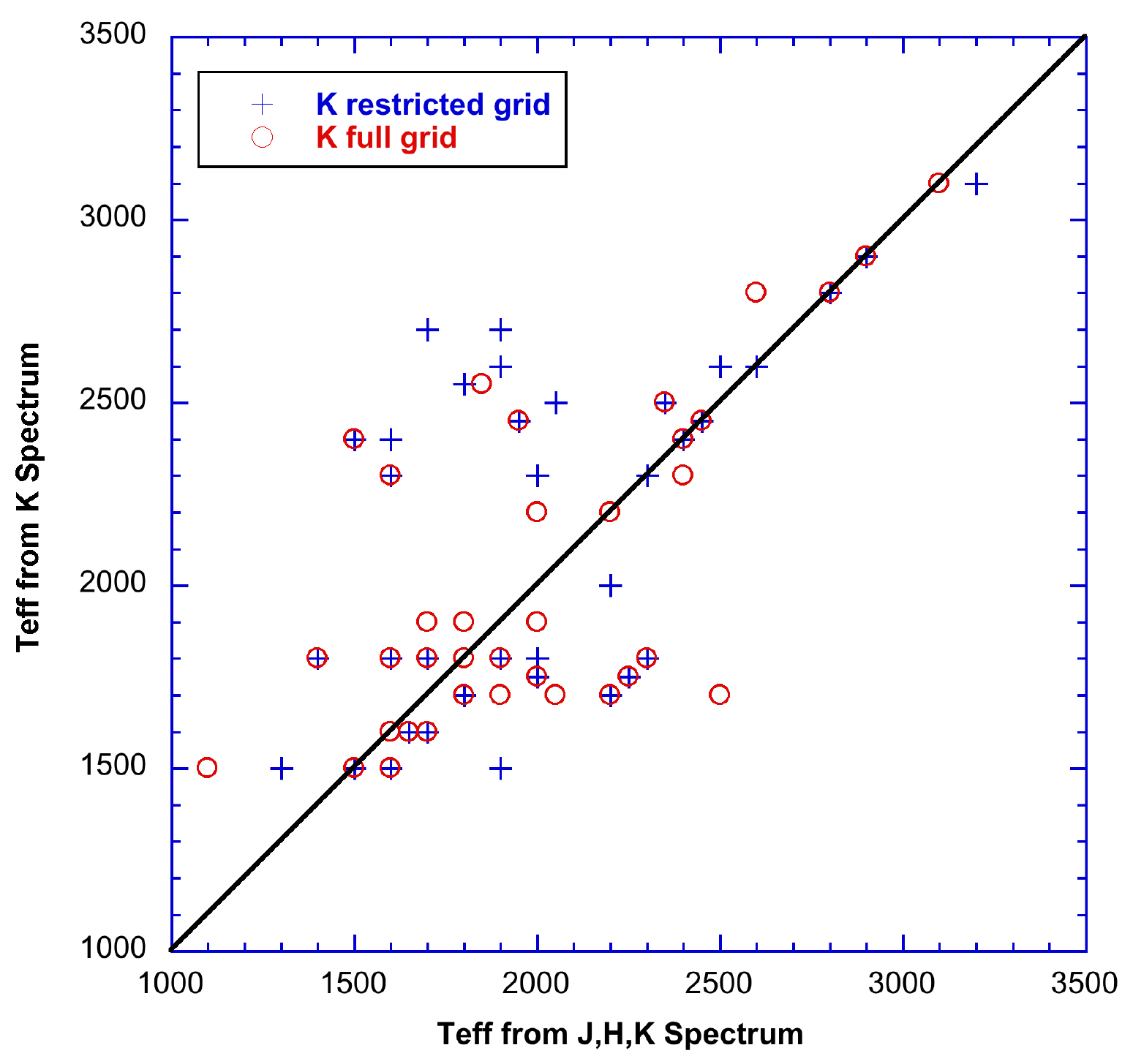}
      \caption{The best-fit effective temperature inferred from a
          single bandpass (left: J-band, middle: H-band, right: K-band)
        portion of the spectrum of each companion as a function
      of the effective temperature based on the full J, H, K 
      spectrum. Results from the full grid are shown with open circles
      and from the restricted log(g) grid with crosses. All five models give systematically higher
      temperatures from fits using only the J-band spectrum.}
         \label{JvsHK}
   \end{figure*}

\subsubsection{Restricted log(g) Model Grid}

Since the targets all have well-constrained young ages, the spectra were also compared with
a restricted model grid limited to log(g) values appropriate for each age range, as detailed
in Section 4 and based on the evolutionary models shown in Figure \ref{EvModels}
\citep{Chabrier:2000}. Further support for using a restricted log(g) grid comes from
measurements of the Na index defined in \citep{Allers:2007} and shown to be sensitive to
surface gravity, but independent of spectral type. For the six targets for which the J-band
data had sufficient sensitivity (2M1207A, TWA 5B, AB Pic B, GSC08047 B, and 2M0141), five
had indices ranging from 0.93 to 1.08, within the range associated with giant stars and
young brown dwarfs in star-forming regions \citep{Allers:2007}. Only 2M0141 has a higher
value of 1.14, though this target may be older than the rest of the sample. The inferred
values of \teff\ from the restricted grid model fits to the data are used to estimate the
range of possible temperatures consistent with the observations. Uncertainties based on
comparisons of the five different synthetic atmospheres should be more representative of the
true range of values than uncertainties estimated from a comparison with a single model set.

The SINFONI data and best-fit atmosphere model (restricted grid) from each of the five
models is shown in Figures \ref{Grid1}, \ref{Grid2}, and \ref{Grid3} for the three age
groups. The region of the spectrum most difficult to fit for the models is the peak and
slope on the blue side of the H-band. The models often show a steeper slope and higher peak
than the data. The peak of the H-band is also shaped by the dust cloud opacity, which is
largely flat across the H-band. A high cloud opacity can flatten the spectrum over this
wavelength range, and may explain the model discrepancy in the H-band shape. The portion of
the data with the lowest signal-to-noise is the blue edge of the J-band, and improving this
section of the data would enable more reliable fits, as some of the targets show a range
of model best fits with significantly different slopes across the J-band. AB Pic B presents
the most unusual shape, with a J-band notably lower flux than the H-band and no model is
able to match the combination of flux levels and spectral shape as well as the other targets
without adding other effects artificially. Unlike 2M1207 B, the inferred radius is not
anomalously low. The inferred radii of all targets is discussed in section 6.2.3.

The results of the model fits to the full and partial spectra of each target are given in
Table \ref{table:restgrid}. Although some of the effective temperatures based on fits to
subsections of the data are different for the full and restricted grids, the \teff\ values
for all targets using the full J,H,K spectrum differ by 100K or less with either the full
grid or the restricted log(g) grid. The exception to this is 2M1207B where the difference in
effective temperature is 200K. The trend of higher inferred temperatures from J-band fits
compared to H+K-band fits is seen in both the full grid fits and the restricted log(g) fits,
as shown in Figure \ref{JvsHK}. The higher values of \teff\ from the J-band fits compared to
the H+K fits are most pronounced for the lower temperature objects below 2200K. There may be
too much dust in the models in this temperature regime, resulting in underestimated J-band
flux which a J-band fit compensates for with a bias towards higher \teff\ values.

\begin{figure}
 \centering
   \resizebox{\hsize}{!}{\includegraphics{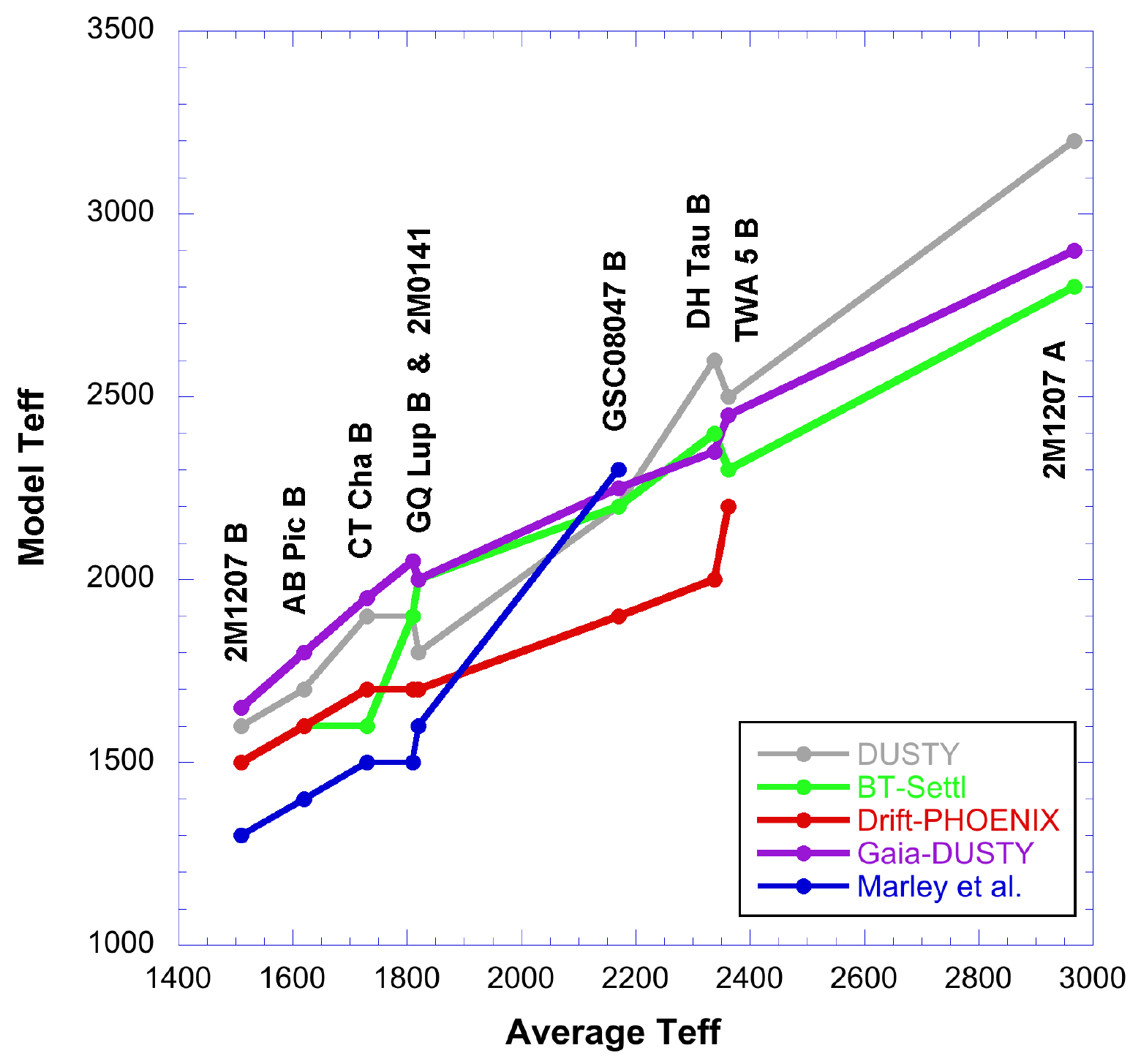}}
 \caption{The best-fit effective temperature of each object
     inferred from the restricted log(g) fits to each
        model (DUSTY - grey, BT-Settl - green, Drift-PHOENIX - red, Gaia-Dusty
        - purple, Marley et al. - blue) plotted against the average
        of all the plotted model
   fits, excluding values at the edge of each grid. Considering all
   models, each object has a range of $\pm$150-300 K, and the
   dispersion in temperature does not show a clear trend as a function
 of average effective temperature.}
 \label{Teff}
\end{figure}

Figure \ref{Teff} shows the inferred \teff\ from each model as a function of the average
\teff\ value. The calculation of the average excluded any values that corresponded to
the limits of a model grid, and those limits are noted in Figure \ref{Teff}. The range in
best-fit temperatures for a given object is $\pm$150-300K, considering all of the five
different models. While no model always suggests the highest or lowest temperature, the
DUSTY and Gaia-Dusty models give typically higher effective temperatures, and the
Drift-PHOENIX and Marley et al. models are usually associated with the cooler temperature
fits for each target, which is consistent with the typically lower amount of dust compared
to the DUSTY models.

  \begin{figure}
   \centering
   \resizebox{\hsize}{!}{\includegraphics{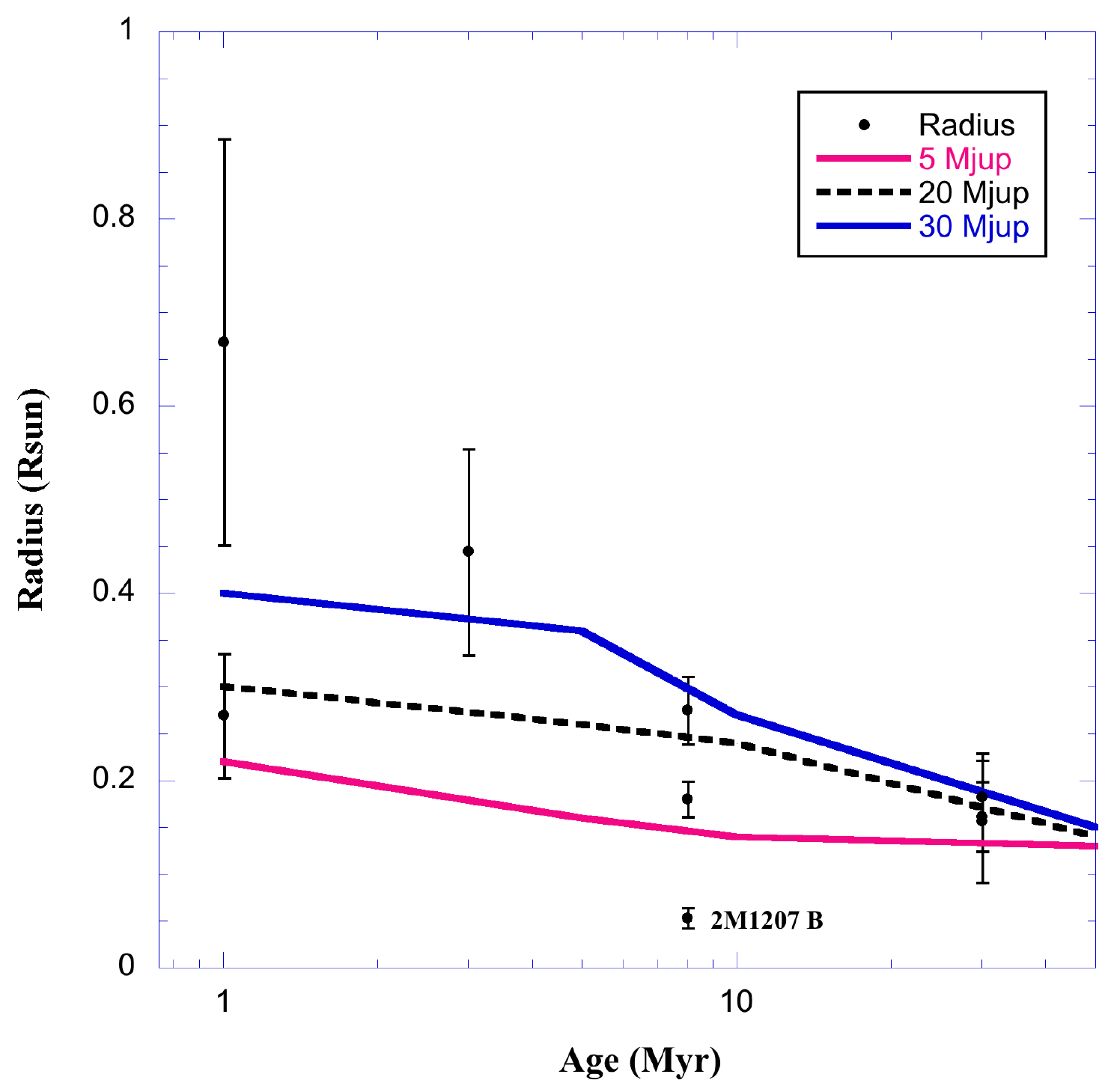}}
      \caption{The radius as a function of age for the targets (filled
      circles) and evolutionary tracks \citep{Baraffe:2003} for
      several substellar objects: 
      5\Mjup\ (pink line), 20\Mjup\ (black dashed line), and
      30\Mjup\ (blue line).}
         \label{radius}
   \end{figure}

\subsubsection{Inferred Radii}

The fitting procedure to obtain the best match for effective temperature from each model
grid includes adjusting the object radius until the flux level corresponds to the synthetic
spectrum. For the restricted log(g) fits reported in Table \ref{table:restgrid}, each model
\teff\ and log(g) combination has an associated object radius. The range of inferred radii
from the five separate models is given in Table \ref{table:radii}; values based on best-fits
at an extreme of a model grid are excluded. Although the radius is estimated from a
restricted log(g) atmosphere model grid, the radius is not fixed by the value of log(g).

As expected for young objects in the process of contracting, the radii are larger than the
$\sim$0.1 \Rsolar\ predicted for older field planets and brown dwarfs \citep{Baraffe:2003}.
The only exception is 2M1207 B, and the apparent small radius -- or underluminosity -- of
this object has been reassessed with updated models different from those considered in this
study and found to be have an inferred radius consistent with  evolutionary models
\citep{Barman:2011}. Due to the young ages of the targets, the radii may retain an imprint
from the initial conditions, including the early accretion history of the objects
\citep[e.g.,][]{Baraffe:2002, Fortney:2008, Baraffe:2010}. Figure \ref{radius} plots the
inferred radii of the targets as a function of age, along with the evolutionary tracks for
the size of 5-30\Mjup\ objects. The data exhibit a systematic decline with stellar age and
are largely within the range predicted by the models \citep{Baraffe:2003}.

\begin{table}
 \centering
 \caption{Inferred Radii}
\label{table:radii}
\begin{tabular}{lc}
 \hline
 \hline
 Target 			& Radius (\Rsolar)		\\
 \hline
  \multicolumn{2}{c}{\bf{1-3 Myr}} \\
 \hline
DH Tau B		& 0.20 -- 0.34  	   \\
GQ Lup B		& 0.45 -- 0.89  \\
 CT Cha B		& 0.34 -- 0.55    \\
 \hline
 \multicolumn{2}{c}{\bf{8 Myr}} \\
 \hline
2M1207 A	& 0.16 -- 0.20  \\
TWA 5B		& 0.24 -- 0.31	   \\
2M1207 B	& 0.043 -- 0.065 	 \\
 \hline
\multicolumn{2}{c}{\bf{30-50 Myr}} \\
 \hline
GSC 08047 B	& 0.12 -- 0.19	   \\
2MASS 0141	& 0.13 -- 0.20	  \\
AB Pic B		& 0.14 -- 0.23  \\
 \hline
\end{tabular}
\end{table}

\begin{table}
 \centering
 \caption{Best Fitting Atmosphere Model Parameters, Metallicity Comparison}
\label{table:metal}
\begin{tabular}{lrrcrl}
 \hline
 \hline
 Wavelength Range 			& \multicolumn{3}{c}{Drift}	& \multicolumn{2}{c}{Drift, [M/H]=0.0}	\\
  					& \teff\ (K) & \logg & [M/H]
                                        & \teff\ (K) & \logg  				\\
 \hline
  \multicolumn{6}{c}{DH Tau B$^a$} \\
 \hline
 JHK (1.10--2.46$\mu$m)		& 2700	& 3.5  & -0.6  & 2000	& 4.0   \\
 J (1.10--1.34$\mu$m)			& 1600	& 3.5  &  0.0  & 1600	& 3.5   \\
 HK (1.46--2.46$\mu$m)		& 2700	& 3.0  & -0.6  & 2700	& 4.0   \\
 \hline
 \multicolumn{6}{c}{GQ Lup B$^a$} \\
 \hline
 JHK (1.10--2.46$\mu$m)		& 1700	& 3.5  &  0.0  & 1700	& 3.5    \\
 J (1.10--1.34$\mu$m)			& 2600	& 3.5  & -0.6  & 2600	& 4.0   \\
 HK (1.46--2.46$\mu$m)		& 1600	& 4.0  & -0.6  & 1700	& 4.0   \\
 \hline
\multicolumn{6}{c}{CT Cha B$^a$} \\
 \hline
 JHK (1.10--2.46$\mu$m)		& 1600	& 3.0  & -0.6  & 1700	& 3.5   \\
 J (1.10--1.34$\mu$m)			& 1700	& 3.0  &  0.0  & 1700	& 3.0   \\
 HK (1.46--2.46$\mu$m)		& 1800	& 4.0  &  0.3  & 1800	& 4.0   \\
 \hline
\multicolumn{6}{c}{2MASS 1207 A} \\
 \hline
 JHK (1.10--2.46$\mu$m)		& 2900	& 4.5  & -0.6  & 3000$^b$	& 4.5    \\
 J (1.10--1.34$\mu$m)			& 2900	& 4.5  & -0.6  & 2900$^b$	& 4.5  \\
 HK (1.46--2.46$\mu$m)		& 2900	& 4.5  & -0.6  & 3000$^b$	& 4.5   \\
 \hline
 \multicolumn{6}{c}{TWA 5 B} \\
 \hline
 JHK (1.10--2.46$\mu$m)		& 2200	& 3.0  &  0.0  & 2200	& 3.0   \\
 J (1.10--1.34$\mu$m)			& 2700	& 4.0  & -0.6  & 2800	& 4.5   \\
 HK (1.46--2.46$\mu$m)		& 1800	& 4.0  &  0.3  & 1700	& 4.5  \\
 \hline
 \multicolumn{6}{c}{2MASS 1207 B} \\
 \hline
 JHK (1.10--2.46$\mu$m)		& 1700	& 4.0  &  0.3  & 1500	& 4.0   \\
 J (1.10--1.34$\mu$m)			& 1500	& 3.5  & -0.6  & 1700	& 4.0   \\
 HK (1.46--2.46$\mu$m)		& 1700	& 4.0  &  0.3  & 1500	& 3.0   \\
 \hline
\multicolumn{6}{c}{GSC 08047 B} \\
 \hline
 JHK (1.10--2.46$\mu$m)		& 1900	& 4.5  &  0.0  & 1900	& 4.5   \\
 J (1.10--1.34$\mu$m)			& 1600	& 3.5  &  0.0  & 1600	& 3.5   \\
 HK (1.46--2.46$\mu$m)		& 1800	& 4.5  &  0.3  & 1700	& 4.0  \\
 \hline
\multicolumn{6}{c}{2MASS 0141} \\
 \hline
 JHK (1.10--2.46$\mu$m)		& 1700	& 4.0  &  0.0  & 1700	& 4.0   \\
 J (1.10--1.34$\mu$m)			& 1600	& 3.5  &  0.0  & 1600	& 3.5   \\
 HK (1.46--2.46$\mu$m)		& 1800	& 5.0  &  0.3  & 1800	& 5.0   \\
 \hline
 \multicolumn{6}{c}{AB Pic B} \\
 \hline
 JHK (1.10--2.46$\mu$m)		& 1600	& 4.5  &  0.0  & 1600	& 4.5   \\
 J (1.10--1.34$\mu$m)			& 1800	& 4.5  &  0.3  & 1600	& 3.5   \\
 HK (1.46--2.46$\mu$m)		& 1800	& 5.0  &  0.0  & 1800	& 5.0   \\
 \hline
\end{tabular}
\\  {\it Note: } $^a$ These star formation region members have been dereddened as described in Sect.\,4.
$^b$ Note that this \teff\ falls at the extremes of the grid used.  
\end{table}

  \begin{figure}
   \centering
   \resizebox{\hsize}{!}{\includegraphics{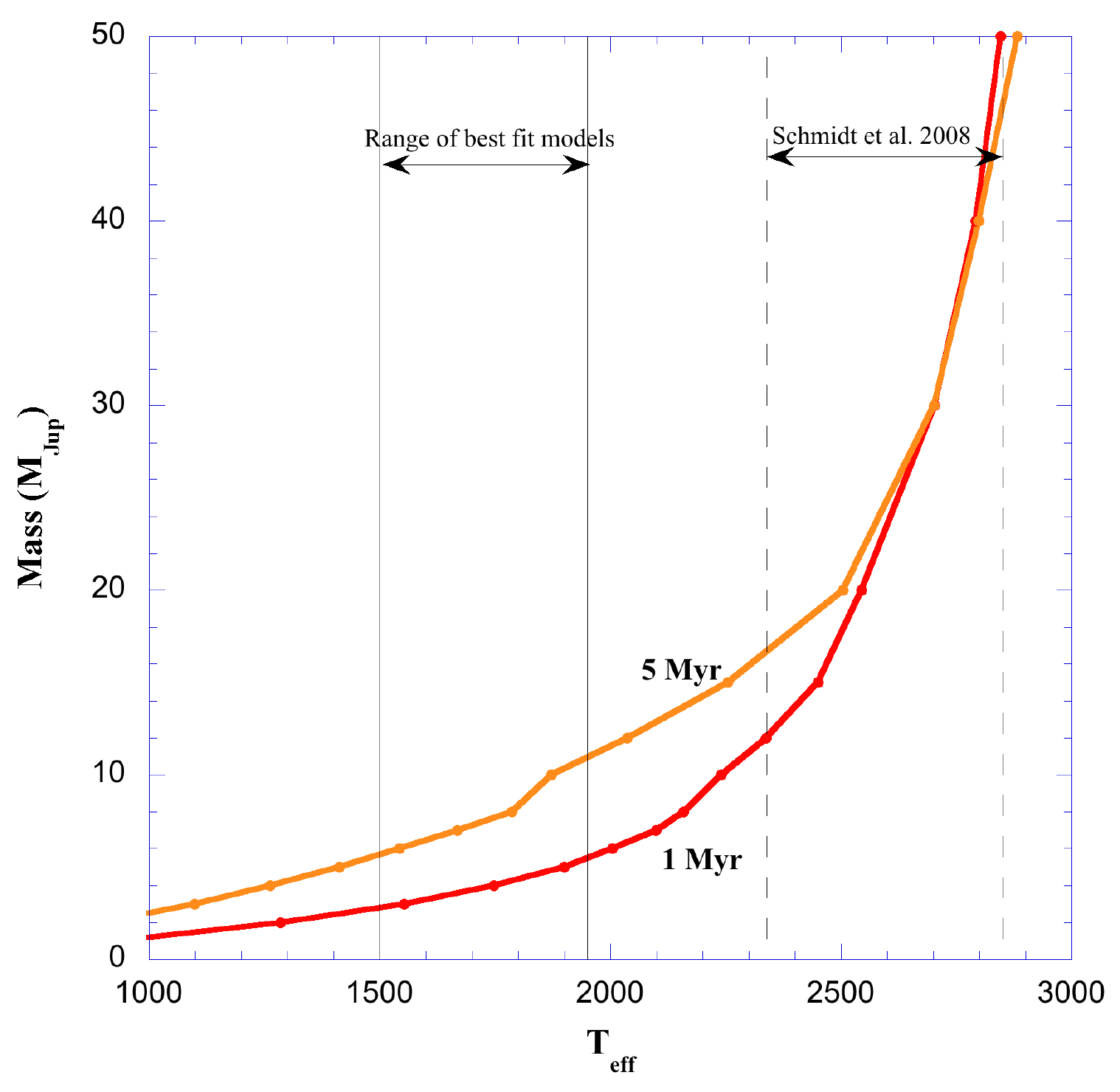}}
      \caption{The mass as a function of \teff\ for substellar
        objects, according to evolutionary models of 1 Myr and 5 Myr
        objects \citep{Chabrier:2000}. The range of effective temperatures
        for CT Cha B
        from the best fit models is shown with solid lines and the
        range previously reported \citep{Schmidt:2008} is shown with dashed lines for
        comparison. The age of CT Cha B is expected to be between 1 Myr
        and 5 Myr.}
         \label{CTCha}
   \end{figure}

  \begin{figure}
   \centering
   \resizebox{\hsize}{!}{\includegraphics{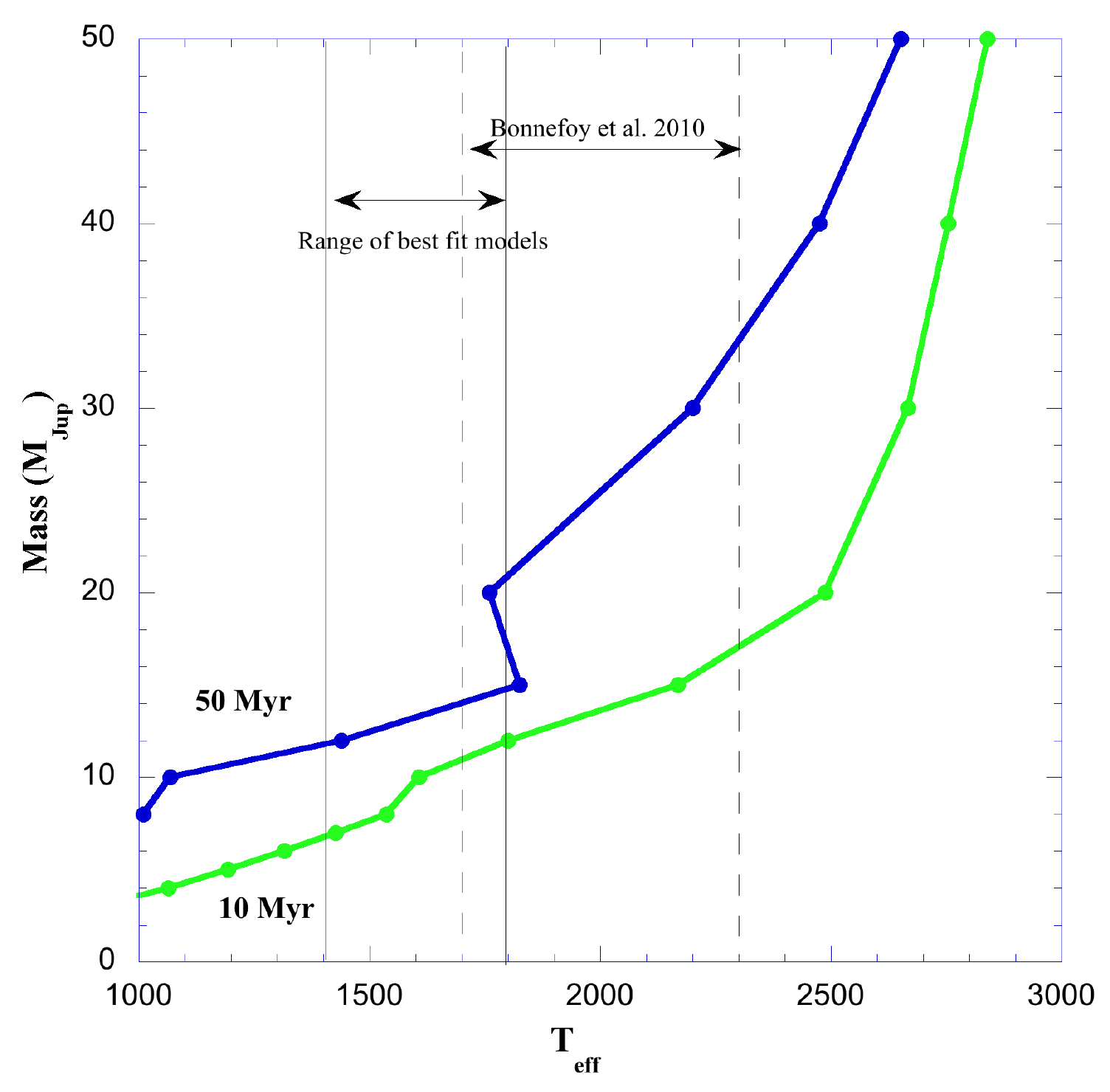}}
      \caption{The mass as a function of \teff\ for substellar
        objects, according to evolutionary models of 10 Myr and 50 Myr
        objects \citep{Chabrier:2000}. The range of effective temperatures
        for AB Pic B
        from the best fit models is shown with solid lines and the
        range previously reported \citep{Bonnefoy:2010} is shown with dashed lines for
        comparison. The age of AB Pic B is expected to be between 10 Myr
        and 50 Myr.}
         \label{ABPic}
   \end{figure}

\subsubsection{Metallicity Grid}

For one of the model grids -- the Drift-PHOENIX set \citep{Witte:2009} -- a range of
non-solar metallicities are also included, and the targets were fit with the expanded
library of synthetic spectra to investigate the impact of metallicity on the inferred
physical properties of \teff\ and log(g). The results of the metallicity comparison are
given in Table \ref{table:metal}, including fits to both the full spectrum and the J- and
H+K-bands. For approximately half the targets, the best fit metallicity over the full
spectrum was the solar value, supporting the use of solar metallicities for these young
stars. For the remaining targets with best fit non-solar metallicities, the impact on the
effective temperature and surface gravity estimates was minimal and resulted in a change of
one step size in log(g) or \teff, except in one case. For DH Tau B, the difference in best
fit effective temperatures is 700K. Overall, the ability to fit for metallicity does not
typically change the inferred properties by an amount larger than the uncertainty estimated
from the range in values from the different models.

\subsection{Comparison with Previous Observations}

\subsubsection{Effective Temperatures}

The inferred effective temperatures for the majority of the sample are consistent with
previous estimates, however two of the targets -- CT Cha B and AB Pic B -- are notably
different. In both cases, the range of best fit values for \teff\ is shifted to cooler
temperatures, with the range of \teff\ 1500-1950 K for CT Cha B compared to 2350-2850 K from
previous estimates \citep{Schmidt:2008} and 1400-1800 K for AB Pic B compared to 1700-2100 K
from the literature \citep{Bonnefoy:2010}. The implications for the masses implied from
evolutionary models \citep{Chabrier:2000} are shown in Figure \ref{CTCha} for CT Cha B and
in Figure \ref{ABPic} for AB Pic B. The lower effective temperatures are associated with
lower masses, overlapping with the planetary mass regime.

The origin of the discrepancy in inferred \teff\ for CT Cha B is clear -- the different
treatment of extinction. By treating A$_{V}$ as a free parameter rather than using the value
measured for the primary, \citet{Schmidt:2008} find a much higher extinction which results
in a large change in slope across the J-K range. For AB Pic B, a previous model comparison
\citep{Bonnefoy:2010} considered each of the J, H and K bands separately, and so the
different results reported here arise from the systematically higher temperatures derived
from fits restricted to the J-band, as shown in Figure \ref{Grid1}. The \teff\ values in
Table \ref{table:restgrid} estimated from the J-band portion of the spectrum are comparable
to the upper bound of the previously reported effective temperature. In both cases,
obtaining spectra over a larger wavelength range will enable more accurate estimates of
\teff\ and determine if CT Cha B and AB Pic B are analogues of 2M1207 B and represent wide
orbit companions with planetary masses, ideal cases for high signal-to-noise measurements of
very low mass companions.

\subsubsection{Variability}

For two of the targets, DH Tau B and 2M0141, there are previously reported J,H,K spectra
taken at different epochs from the SINFONI data, and the combination of data sets are used
to test for variability in the spectral shape, such as observed for the young low mass
object TWA 30 in the TW Hydra Association \citep{Looper:2010}. DH Tau B and 2M0141 are
also young objects bracketing the age of TWA 30, making these targets
interesting comparisons to TWA 30. The
first epoch of DH Tau B spectra were obtained in 2003 (K-band) and 2004 (J+H-band) with
CISCO at Subaru \citep{Itoh:2005}, several years prior to the SINFONI observations in 2007
(Table \ref{table:obs}). Over the 3-4 year time period between observations, the spectrum
shape does not change. The first epoch of the 2M0141 observations were obtained in 2004 with
Spex on the IRTF \citep{Kirkpatrick:2006}, two years prior to the 2006 SINFONI observations
(Table \ref{table:obs}). Like DH Tau B, no variability in the spectral
shape is seen. It is not possible to search for a grey brightening or
dimming of the spectrum, such as that reported in Apai et al. (2012, submitted)
because the flux calibration of the SINFONI data relies on the same photometry used to calibrate the
first epoch spectra,. Additional coordinated
photometric and spectroscopic measurements are required to determine the frequency,
wavelength dependence and physical origin of variability in low mass object atmospheres.

\section{Summary and Conclusions}

With data from the SINFONI integral field unit, an empirical grid of J,H,K spectra for young
$\sim$1-50 Myr substellar objects has been constructed and compared with a suite of five
theoretical atmospheric model grids -- DUSTY \citep{Allard:2001}, BT-Settl
\citep{Allard:2003, Allard:2010}, Drift-PHOENIX \citep{Helling:2008b,Witte:2009}, Gaia-Dusty
\citep{Rice:2010}, and Marley et al. \citep{Ackerman:2001, Marley:2003, Saumon:2008,
Freedman:2008}. The spectra should also serve as observational comparisons for updated
models including more advanced treatment of physical processes. Since most of these objects
are companions, there is additional information on the age and extinction of the substellar
object to aid in the model fitting. All spectra show the distinct triangular shape over the
H-band, a region of the spectrum which was found to be particularly difficult for most
models to match in detail. 

Based on the best-fit synthetic atmosphere spectrum for each target from each 
restricted log(g) model grid, the uncertainty in the temperature estimate for a young,
substellar object is $\pm$150--300 K from fits to the J,H,K spectrum. This is considerably
larger than the uncertainty usually inferred from comparisons to a single model grid
\citep[$\sim$ 50--100\,K, cf.][]{Burrows:2006}. For all targets except 2M1207B, the inferred
radius from the fit is greater than 0.1\Rsolar, and generally consistent with the
expectations of evolutionary models of young brown dwarfs
\citep{Baraffe:2003}. 

We note that fits to small wavelength ranges do not account for the broad spectral
morphology and so may suggest effective temperatures incompatible with observations of a
broader range in wavelength. Fitting only
the J-band portion results in systematically higher temperatures than fitting the H+K-band,
likely due to systematic uncertainties in the model dust opacities. When the spectrum covers
the full J,H,K range, the inferred temperatures are approximately the same whether or not
the model grids include surface gravities limited to the range for young objects or expanded
to include the higher values of older objects.  While the temperatures were consistent
using the full or restricted log(g) model grids, it is important to note that the spectral
shape fit to the full model grid did not find the expected low surface gravities of these
objects with known young ages. This apparent inconsistency is due to
the fact that differences in spectral morphology caused by different
effective temperatures dominate over spectral changes due to low surface gravity. Empirical
comparison with observed spectra such as these may provide a more reliable method to
determine the surface gravity; specific spectral lines and indices will be investigated in a
follow-up paper (King et al. 2011). For the Drift-PHOENIX model grid, it was possible to
include fits to non-solar metallicities, and varying the metallicity did not impact the
inferred temperatures by more than the uncertainty estimated from the range of model fits
presented here.

From comparison of the SINFONI data with previous spectra of two targets, there is no
evidence of substantial changes in the spectral shape, however more observations and
coordinated photometry and spectroscopy are required to rule out variability in these young
substellar objects. The spectrum of AB Pic B shows an unusually low J-band spectrum relative
to the H-band spectrum, analogous to 2M1207B, and the overall shape of the spectrum is more
difficult to fit with the model grid than the remaining targets. For CT Cha B and AB Pic B,
the new fits to the spectra suggest that the objects may have lower temperatures and,
consequently, lower masses compared to previous estimates \citep{Schmidt:2008,
Bonnefoy:2010}. Additional wavelength coverage in the spectra will refine the temperature
estimates to determine if these companions represent additional examples of planetary mass
companions that can be investigated with high signal-to-noise spectroscopy, like 2M1207 B.

\begin{acknowledgements}
We gratefully acknowledge grant support to Exeter from the Leverhulme Trust (F/00144/BJ) for
J. P. and R. R. K., and STFC (ST/F003277/1) for J. P. and A. V. The studentship for R. DR.
is provided by STFC (ST/F007124/1). ChH acknowledges an ERC starting grant from the the EU program FP7 {\it Ideas}
for the LEAP project. We thank F. Allard, T. Barman, M. Marley, and D. Saumon
for constructive comments and providing theoretical model grids used
in this analysis, and we thank the referee for suggestions that
improved the paper.

\end{acknowledgements}

\bibliographystyle{aa}
\bibliography{bib2}

\end{document}